\documentclass[%
 reprint,
 amsmath,amssymb,
 aps,
 pre,
]{revtex4-1}

\usepackage{graphicx}  
\usepackage{dcolumn}   
\usepackage{bm}        
\usepackage{amssymb}   
\usepackage{mathtools}
\usepackage{amsmath}
\usepackage{color}

\newcommand{\beginsupplement}{%
       \setcounter{table}{0}
       \renewcommand{\thetable}{S\arabic{table}}%
       \setcounter{figure}{0}
       \renewcommand{\thefigure}{S\arabic{figure}}%
       \setcounter{subsection}{0}
       \renewcommand{\thesubsection}{S\arabic{subsection}}%
    }

\usepackage{xr}
    
\begin{document}

\title{Geometric signatures of tissue surface tension in a three-dimensional model of confluent tissue }

\author{Preeti Sahu\textit{$^{1,2,*}$}, J. M. Schwarz\textit{$^{1,3}$} and M. Lisa Manning\textit{$^{1,}$}}
\email{preeti.sahu@ist.ac.at; mmanning@syr.edu }
\affiliation{$^1$ Department of Physics and BioInspired Syracuse: Institute for Material and Living Systems, Syracuse University, Syracuse, NY 13244, USA\\
$^2$ IST Austria, Am Campus 1, 3400 Klosterneuburg, Austria\\
$^3$ Indian Creek Farm, Ithaca, NY 14850, USA}

\date{\today}

\begin{abstract}
In dense biological tissues, cell types performing different roles remain segregated by maintaining sharp interfaces. To better understand the mechanisms for such sharp compartmentalization, we study the effect of an imposed heterotypic tension at the interface between two distinct cell types in a fully 3D model for confluent tissues. We find that cells rapidly sort and self-organize to generate a tissue-scale interface between cell types, and cells adjacent to this interface exhibit signature geometric features including nematic-like ordering, bimodal facet areas, and registration, or alignment, of cell centers on either side of the two-tissue interface. The magnitude of these features scales directly with the magnitude of imposed tension, suggesting that biologists can estimate the magnitude of tissue surface tension between two tissue types simply by segmenting a 3D tissue. To uncover the underlying physical mechanisms driving these geometric features, we develop two minimal, ordered models using two different underlying lattices that identify an energetic competition between bulk cell shapes and tissue interface area. When the interface area dominates, changes to neighbor topology are costly and occur less frequently, which generates the observed geometric features.
\end{abstract}

\pacs{}
\maketitle

\section*{\label{intro}Introduction}
An important collective phenomenon observed in groups of biological cells is the process of cell sorting, where cells of different types spontaneously spatially segregate into separate compartments.  These distinct compartments not only play an integral role during an organism’s formative stages~\cite{Trinkaus1955a,Waites2017b,Unbekandt2010,Montero2005,Klopper2010}, but are also crucial for the upkeep of normal functioning of organs~\cite{Rubsam2017,Cochet-Escartin2017a} and for containment of the spread of diseased/infected tissues~\cite{Friedl2004,Foty2005,Pawlizak2015,Song2016}. Broadly speaking, cell sorting mechanisms can be classified into two generic  categories- (a) biochemical/morphogen gradients~\cite{Turing1990,Streichan2018} and (b) differences in mechanical properties of individual cells. These mechanical properties can include cell-cell adhesivity~\cite{Steinberg1963a,STEINBERG2007281}, acto-myosin contractility~\cite{Harris1976b,Brodland2002a,Mertz2012a,Maitre2012b}, a mechano-chemical coupling between both cell-cell adhesivity and acto-myosin contractility~\cite{Amack2012a,Manning2010c,Engl2014}, or an explicit interfacial tension between unlike cells, often called heterotypic interfacial tension (HIT)~\cite{Graner1992,Barton2017,Canty2017a,Sussman2018b}. 

Much of the computational and theoretical work on cell sorting has focused on particle mixture simulations. In such mixtures, the mechanism for sorting relies heavily on active fluctuations~\cite{PhysRevE.85.031907,Merks2005,Sun2013,Jiang1998,Palsson2008}.  
However, an essential feature of cell sorting that is observed in experimental co-cultures is that the interface is much sharper than what is expected from a particulate mixture~\cite{Landsberg2009a,Manning2010b,Dahmann2011,Nnetu2012a,Monier2011,Calzolari2014}. While such a straight and sharp interface is difficult to obtain merely by diffusive morphogens, heterotypic interfacial tensions in confluent tissues, where there are no gaps between cells, can generate a sharp interface easily~\cite{Kesavan2020,Sussman2018b,Sahu2020b}. 
In such cases, the fact that a confluent monolayer must tessellate space, which is captured in vertex or Voronoi models for tissues, results in forces that are discontinuous functions of cell displacements. It is precisely this non-analytic behaviour resulting from topological interactions between cells that drives sharpening in two dimensions. And yet, the tissue remains fluid-like such that the sharp interface is also easily deformable. Such sharp but deformable interfaces are not observed in particle-based models with metric interactions between cells. 

While confluent monolayers, with a single layer of cells, are biologically relevant, fully three-dimensional confluent tissues, such as stratified epithelia or early vertebrate embryos, are even more ubiquitous. Therefore, it is important to determine if/how the topological nature of the interactions between the cells also drive sharp but deformable interfaces in three dimensions. Prior work on topological models for cell sorting in three dimensions focused on cells coalescing into small clusters via a Rayleigh-Plateau instability as well as regions of mixed cell types untangling to facilitate compartmentalization in the absence of fluctuations~\cite{PhysRevLett.101.148105}. In this manuscript we instead focus on quantifying the dynamics and cell geometries in the presence of fluctuations, and identifying the topological mechanisms driving cell sorting in three dimensions. 

One goal is to study the dynamics of cell sorting in confluent fluid-like tissues by implementing HIT in the presence of fluctuating forces in 3D. Indeed, we demonstrate that HIT is an efficient sorting mechanism and that, unsurprisingly, the magnitude of HIT governs the timescale for segregation in 3D.

Unfortunately, it is rather difficult to test this prediction in experiments, as it is quite difficult to measure the magnitude of HIT. In laser ablation experiments on monolayers, cuts are made to ablate a cell-cell junction with the help of a pulsed laser. The resultant relaxation dynamics can help determine line tensions~\cite{Sugimura2016}. However, ablating interfaces and recording retractions along arbitrary interfaces is very difficult in 3D, and has not been yet used to estimate HIT in 3D. Indirect measures using few-cell assays, such as double pipette aspiration experiments, suggest that HIT can create robust cell sorting in 3D tissues~\cite{Rubsam2017,Tsai2019,Maitre2012b}. 

However, using isolated cells for measuring effective tension may not provide a complete picture as a confluent neighborhood can significantly change a cell's mechanics~\cite{Canty2017a,Sussman2018b}. Recent work has shown that geometrical properties of interfacial cells in a confluent neighborhood can be directly affected by increased interfacial contractility and tension~\cite{Okuda2015,Sussman2018b,Yang2009}. Therefore, a second goal is to explore the idea that cellular geometry can perhaps be used as a simpler and more direct readout of HIT in 3D mixtures, as geometric features have recently become more accessible in experiments due to advances in tissue segmentation techniques.

In addition to developing tools for measuring HIT, a third goal of this work is to identify the mechanisms driving cell sorting in 3D. For a fluid-like particulate mixture, the mechanism is simple, as the minimum energy state is a configuration that minimizes the shared surface area. In the presence of large enough fluctuations the system can find a simple geometry that achieves this goal via complete spatial segregation with a strong gradient at the interface~\cite{CanongiaLopes2002,Majumder2011,Chremos2014}. In confluent tissues, however, there are cell-scale geometric constraints that may compete with macroscopic interfacial dynamics. Specifically, in vertex models for isotropic confluent tissues~\cite{Nagai2001, Farhadifar2007a,Teleman2007,Staple2010a,Manning2010b,Hilgenfeldt2008, Chiou2012,Bi2015c,Fletcher2014,Merkel2018}, cells attempt to attain a preferred cell shape index, which is dimensionless ratio of perimeter P and area A i.e. $ s_0^{2D}  = P/\sqrt{A}$ in 2D~\cite{Bi2015c,Bi2016b}, and of surface area S and volume V i.e. $s_0^{3D} = S/V^{2/3}$ in 3D~\cite{Merkel2018}. If the cells are able to attain their preferred cell shape, which generally happens in isotropic tissues for elongated shapes with $s_0^{2D} >~ 3.81 $ and  $s_0^{3D} >~ 5.41 $, then the tissue is fluid-like, while if they are unable to attain that shape the tissue is solid-like. 

Moreover, as discussed above, 2D results suggest specific features of the 2D cell-scale geometry help pin cells at a heterotypic interface, although it is not trivial to generalize the arguments to 3D. Therefore, we study whether these cell-scale constraints, which vary with tissue rheology, impact cell sorting in 3D. We develop simple toy models to demonstrate that a competition between the bulk cell shape preference and geometric pinning of cell shapes at the heterotypic interface drive both cell sorting and the formation of specialized geometric features at the HIT interface. Finally, we confirm the predictions made by the toy models with full numerical simulations.

\section*{Model}

To understand the interfacial mechanics between two cell types, we use the recently developed 3D Voronoi model~\cite{Merkel2018}. A system with periodic boundaries is created using a Voronoi tessellation of the cell centers of $N$ cells. Individual cells  have preferred volumes $V_0$ and surface area $S_0$. The combination of volume incompressibility and surface area regulation due to adhesion and contractility generate a preferred cell shape index $s_0=S_0/ V_0^{2/3}$. For example, a regular BCC unit cell (truncated octahedron), has a dimensionless shape index of $s_0\sim 5.31$~\cite{Lucarini2009}. Here we set $V_0$ to 1. Half of the cells are tagged differently, creating a mixture of two cell types- $\beta=1$ or $\beta=2$, which are otherwise identical except there is heterotypic interface between cells of different type. Of course, in experimental systems there are likely mechanical differences between different cell types in addition to HIT. Here we ignore those differences to study effects driven by HIT, as our previous work in 2D suggests that sorting driven by innate mechanical differences is significantly weaker than HIT\cite{Sahu2020a}. In addition to the original monodisperse energy functional, we impose an additional surface tension along the heterotypic interface. Therefore cells minimize their mechanical energy using the following dimensionless energy functional:

\begin{equation}
\label{eq:HST_3}
	e = \sum_i{\bigg[k_v(v_i-1)^2+(s_i-s_0)^2\bigg]} +  \sum_{\langle i,j\rangle}(1-\delta_{\alpha\beta})\sigma a_{ij}\text{,}
\end{equation}
where $v_i$ denotes the $i$th cell volume and $s_i$ denotes its surface area, non-dimensionalized by $V_0$. The unit of length is defined such that the average cell volume $\langle V_i\rangle$ is 1. Additionally, $k_s=K_{V}/K_{S}$ sets the ratio between volume and area stiffnesses, and is also set to unity. The second summation imposes an additional surface tension between heterotypic cells, where the sum is over all facets with area  $a_{ij}$ shared between cells $i$ and $j$ of types $\alpha$ and $\beta$ respectively. The surface tension $\sigma$, for simplicity, is assumed to be the same for all facets. It is non-dimensionalized by $K_S V_0^2$ which is unity for our system.  Biological cells can establish heterotypic tension by co-regulating the acto-myosin network and adhesion molecules~\cite{Amack2012a}. A biologically relevant estimate of the heterotypic to homotypic tension ratio, based on examination of the contact angles at cell vertices in ectoderm-mesoderm co-cultures in Xenopus~\cite{Canty2017a}, indicates $\sigma\sim2$ in natural units of the system. 
For systems with fluctuations, we analyse the dynamics of over-damped self-propelled particles with a high angular noise, which effectively leads to Brownian dynamics at the timescales relevant to us. The timescales are reported in units of the self-diffusivity timescale $\tau_s^0$, details of which are provided in Supplemental section \ref{3dspp}. While there are other possible dynamical rules, recent work on 2D mixtures has shown that the properties of an interface between two cells types with HIT between them is rather robust to the specifics of the dynamical rules~\cite{Sussman2018b}. For analyzing behavior in the absence of fluctuations, we use a conjugate gradient minimizer. See Supplemental section \ref{3dspp} for more details.

\section*{Results}
\begin{figure}[htp]
\centering
\includegraphics[width=\columnwidth]{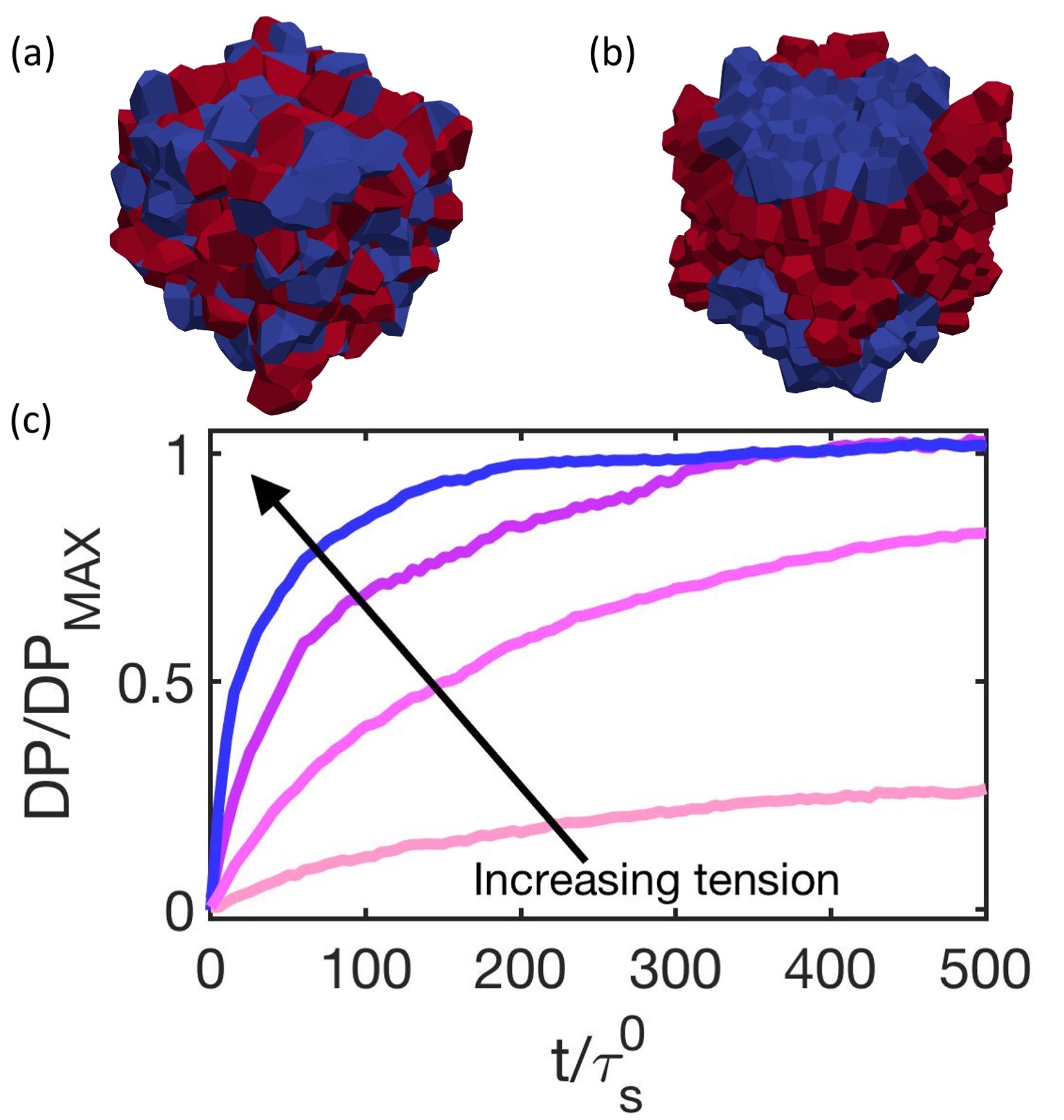}
\caption{\textbf{ Sorting in fluid-like binary tissue with HIT}: Initial(a) and final(b) snapshots of a $s_0=5.5$ mixture with high tension of $\sigma=1$ and system size of $N=512$. Both cell types denoted by red and blue polyhedra. (c) Quantification of segregation: demixing parameter DP versus simulation time in units of self-diffusivity timescale $\tau_s^0$ for increasing tension $\sigma=0.001,0.003,0.010,0.032$ (pink to blue). 
}
\label{fig:dp_3}
\end{figure}

{\it Demixing parameter:} To test if HIT leads to significant segregation in 3D tissues, in a manner similar to that of 2D mixtures~\cite{Sahu2020a}, we first focus on a fluid-like parameter regime~\cite{Merkel2018} where cells undergo diffusive motion ($s_0>5.41$). For a fixed shape index $s_0=5.5$, we start from an initially mixed configuration Fig~\ref{fig:dp_3}(a) with a system size of $N=512$ cells. We let the system evolve long enough so each cell on average explores a distance equivalent to the simulation box length.

Let us now understand the role of HIT on the bulk demixing. For a fixed $\sigma=1$, a final configuration for such a mixture is shown in Fig~\ref{fig:dp_3}(b). It is clearly segregated as compared to the initial snapshot. Some fraction of ensembles are able to create a planar interface as well. To quantify this demixing, we study the demixing parameter DP~\cite{Sahu2020a}, which measures the average neighborhood composition of every cell.
Defining $N_s$ as the number of homotypic (similar cell type) neighbors and $N_t$ as the total number of neighbors, 
\begin{equation}
 \label{eq:dp_3}
DP=\big \langle DP_i \big \rangle = \bigg \langle 2\bigg(\frac{N_s}{N_t} - \frac{1}{2} \bigg) \bigg \rangle,
\end{equation}
where the brackets denote averaging over all cells in the tissue. 
In a completely mixed state, $DP=0$, whereas in a completely sorted mixture, $DP=1$, in the limit of infinite system size. However, as large system sizes can be time-consuming, we compute the maximum attainable value of demixing ($DP_{max}$) for a particular system size by looking at minimal surface configurations as shown in  Fig.~\ref{fig:dp_max}. Hence we plot the demixing parameter relative to this maximum value in Fig~\ref{fig:dp_3}. The value of demixing is zero at the beginning as both the cell types are seeded at random positions, but it soon attains a high value, very close to $DP_{max}$.
In the presence of HIT, the value of demixing increases quickly, indicating that it can efficiently create robust segregation \textemdash very similar to a liquid-liquid particulate mixture and 2D confluent mixtures.

In the presence of fluctuations, we find that HIT efficiently leads to significant segregation in mixed 3D tissues. We also observe that with higher values of tension, the initial phase of the sorting process becomes faster as shown in Fig.~\ref{fig:dp_3}. This confirms that heterotypic surface tension is very effective at compartmentalization in 3D, as expected. 

{\it Geometric features:}  While biological cells are capable of upregulating tension cables along heterotypic interfaces via biochemical pathways, it is very difficult to directly measure this tension within a 3D tissue. Can features of individual cells at the interface help us quantify such tensions? In this section, we focus again on fluid-like systems with $s_0=5.5$.
A visual inspection of the segregated mixture shown in Fig~\ref{fig:dp_3}(b) indicates that the interfacial cells may be more elongated and nematically ordered as compared to the cells in the interior. This observation hints at a direct relationship between the applied tension and the surrounding cellular geometry. To delve deeper into the ways in which the surrounding cells deform, we set up a maximally segregated mixture. Here, both compartments are placed side by side, similar to previous work by Sussman \textit{et al}~\cite{Sussman2018b}. We then study the cellular geometry as a function of the applied interfacial tension. Shape elongation along with the prominent stacking of cells (alignment of the polyhedral long axes in Fig.~\ref{fig:bilayer_panel}(a)) can be observed here as well. We next quantify such geometric effects. 

The first quantity is the steady-state cell shape index $s$. This helps us quantify whether or not the otherwise homogeneous cells remain homogeneous after HIT is established. We measure the shape of both interfacial cells- cells that directly touch the boundary ($s_{boundary}$), and the interior cells ($s_{bulk}$). This further helps isolate the shape changes in the immediate neighborhood of the interface. Individual cells have a final volume $V_i$ and surface area $S_i$. Hence, $s_{boundary}$ is defined as:
\begin{equation}
 \label{eq:s_3}
s_{boundary}= \bigg \langle \frac{S}{V^{2/3}}\bigg \rangle_{boundary},
\end{equation}
and $s_{bulk}$ is similarly averaged over the interior cells. In the absence of HIT ($\sigma=0$), both shapes have the same value of $s_0=5.5$.

Both the shape indices are shown as a function of increasing tension in Fig.~\ref{fig:bilayer_panel}(b). For small values of tension, the shapes indices are very similar at the beginning, but they gradually saturate at a higher value of disparity. In other words, with a higher interfacial tension, the neighboring cells become more elongated, whereas the interior cells become more compact/round.

To study the alignment between cells, we next measure the orientation of interfacial polyhedra. We define orientation vector of a cell as the major axis of its moment of inertia tensor. We then plot the angular distribution of the angle made by each vector with respect to the normal to the interface ($\theta$), which in this bilayer arrangement is simply the z axis. For a homogeneous system with no HIT, the angles are very close to the random distribution density function in 3D, which is proportional to $\sin \theta$. But with a slight increase in tension, the cells polarize and orient themselves perpendicular to the interface, as shown in Fig.~\ref{fig:bilayer_panel}(c). 

Lastly, we study polygonal faces that make up the heterotypic interface by plotting the facet area distribution with respect to increasing tension. This is in analogy to the measurement of edge lengths in 2D work~\cite{Sussman2018b}. With no HIT, the distribution is roughly uniform up to a characteristic length scale, whereas, with increasing tension it becomes bimodal, as shown in Fig.~\ref{fig:bilayer_panel}(d). This means that the facets are either large or vanishingly small at high tensions. With the help of smaller facets, the interfacial vertices come very close to having a vertex coordination higher than the normal tetrahedral coordination, similar to the 4-fold vertices observed along 2D tension cables. The average area of a facet, $<a>$, also increased with increasing tension as shown in Fig.~\ref{fig:facet_max}. This quantitatively confirms that HIT indeed affects the geometry of the surrounding cells, inducing shape changes and nematic-like ordering in an otherwise homogeneous collection of cells.

Since some similar geometric features have been observed in 2D models for confluent tissues~\cite{Sussman2018b}, we hypothesize that the origin of these signatures in 3D might be based on topological interactions between the cells which have been implicated in 2D.

\begin{figure}[htp]
\centering
\includegraphics[width=0.9\columnwidth]{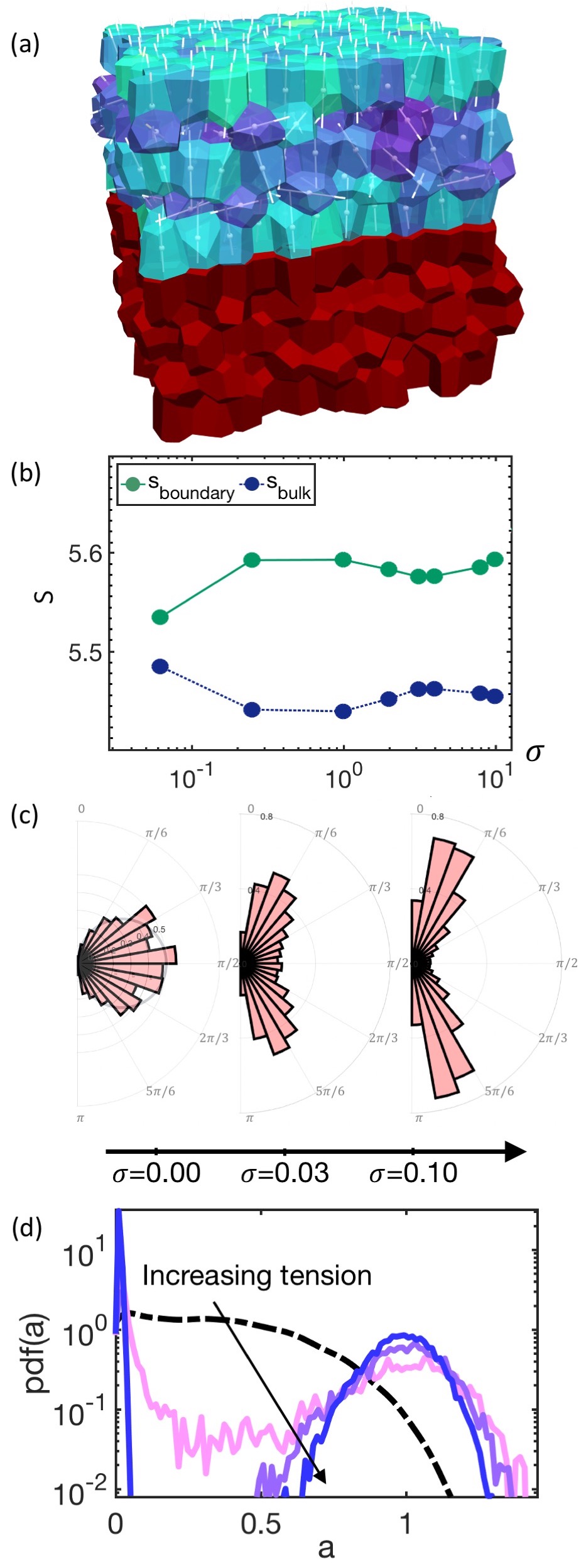}
\caption{\textbf{Cellular geometry changes around the high-tension interface}: (a) Snapshot of the bilayer arrangement of sorted compartments for a high value of HIT, $\sigma=20$. Only the blue cell type is emphasized here, colored by major axis length. Greenish-blue is for elongated and purple for rounder cells. The white rods denote the long axis of the polyhedron. (Caption continued on next page.)}
\label{fig:bilayer_panel}
\end{figure}
\addtocounter{figure}{-1}
\begin{figure} [t!]
  \caption{(Continued from previous page.) (b) Acquired cell shape index (s), plotted with respect to tension ($\sigma$), is higher for interfacial cells (solid green curve) as compared to the cells in bulk (dashed blue curve). (c) Rose plot for the distribution of orientation angle of interfacial cells is shown for increasing tension ($\sigma$). The control distribution ($\sigma$=0) is superimposed on a faint black curve that represents the density for uniform distribution i.e $\sin \theta$. (d) The probability distribution $P(a)$ of heterotypic facet area $a$ is shown for increasing tension $\sigma=0.32,1.00,2.00$ (magenta to blue). For these data, N=512 and $s_0 = 5.5$. }
\end{figure}

In order to determine the mechanism driving these geometric changes, we first focus on the specific geometry of an interfacial cell and ask: what does it take for the system to attain this precise geometry?

\begin{figure}[htp]
\centering
\includegraphics[width=0.95\columnwidth]{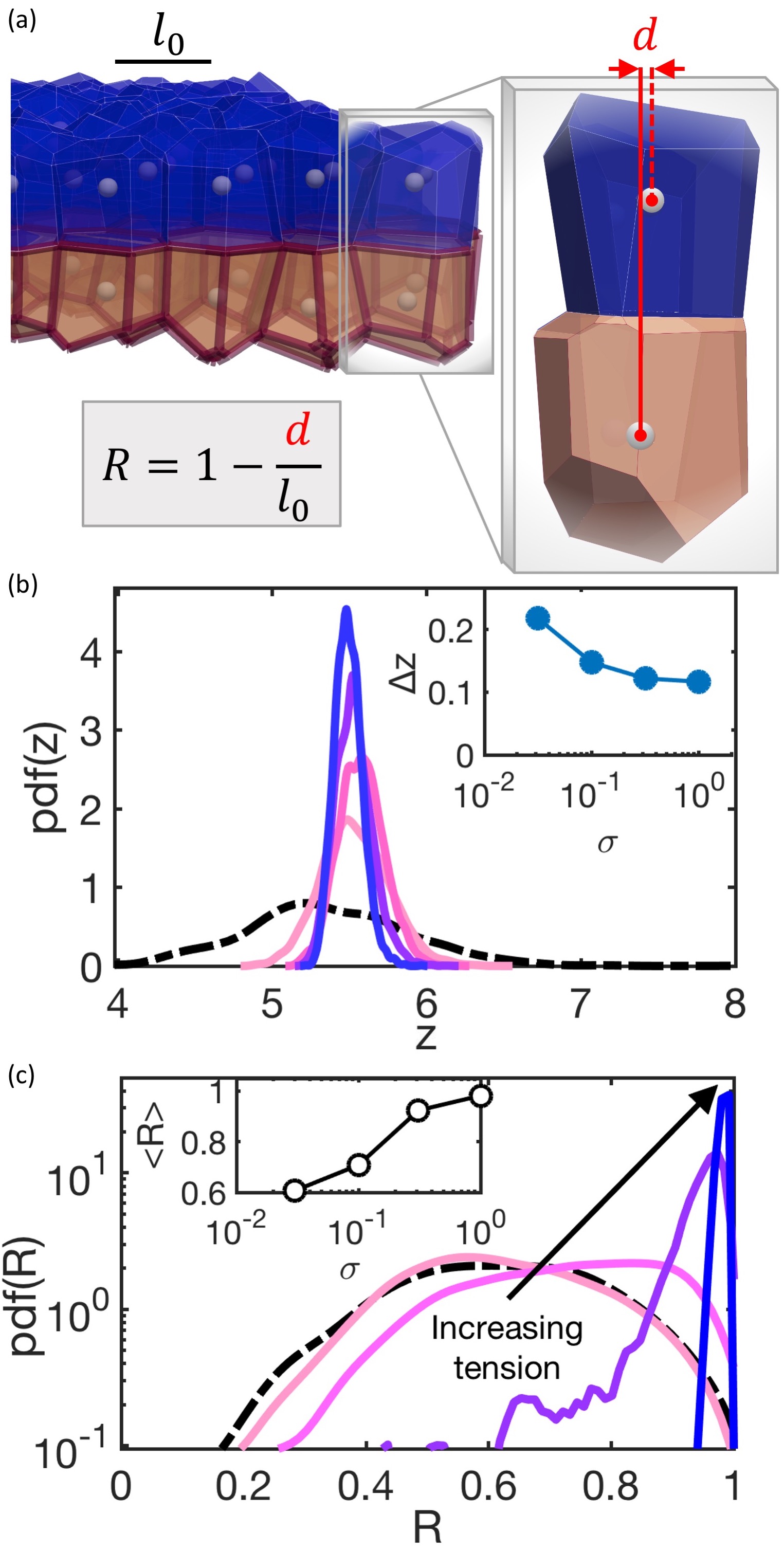}
\caption{\textbf{Interfacial cells are prism-like}: (a) A simulation snapshot highlighting the cells at both sides of the interface. Cell types are tagged in blue and red and made translucent to make their centers (white solid spheres) visible. To characterize the high alignment between the heterotypic cell centers we use cell-cell registration $R$ defined in the main text in terms of the distance $d$ between the centers along the interface (highlighted in red) and the characteristic cell length $l_0$ (denoted by the black scale bar). (b) Distribution $pdf(z)$ of the heights of cell centers $z$ for blue interface cells is plotted for increasing tension $\sigma=0.03,0.10,0.31,1.00$ (magenta to blue). The dashed black line depicts a system with no HIT. The inset shows the standard deviation $\Delta z $ with as a function of the HIT $\sigma$. (c) Distribution of registries pdf($R$) between heterotypic cells is plotted for increasing tension. The inset shows the averaged registry with respect to tension. The lowest value of tension has the same average distance as no tension at all. The circle radii corresponds to the deviation from the mean values.  }
\label{fig:interface_zoom}
\end{figure}

A typical interfacial neighborhood is shown in Fig.~\ref{fig:interface_zoom}(a). A right prism can be defined as a polyhedron with flat top and bottom facets, and perpendicularly aligned lateral facets. The cells here seem to closely resemble the geometry of a one-sided right prism, with the flat side at the interface. The unique shape can be attained in a Voronoi tessellation only by fulfilling two conditions : (a) cell heights are arranged in a plane, and (b) interfacial pairs align their centers so the distance between centers in the XY plane is minimized. One way to quantify this alignment is to measure it as the registration- $R$ between the cell centers. We define it as-
\begin{equation}
\label{eq:R}
R= 1-\frac{d}{l_0} ,
\end{equation}

where $l_0$ is the typical lengthscale and $d$ is the distance between centers along the interfacial plane as depicted in Fig.~\ref{fig:interface_zoom}(a). The value of $l_0$ is set to unity as $V_0^{1/3}=1$ in our systems.
When cell centers are completely registered its value attains a maximum value of unity. 
We quantify both of the aforementioned criteria and find them both fulfilled for higher values of tension, as shown in Fig.~\ref{fig:interface_zoom}(b-c). Moreover, the insets to each figure show that the cell geometries approach the right-prism condition monotonically as a function of increasing HIT.

From work in 2D we understand that perturbations perpendicular to the surface of the interface are very costly~\cite{Sussman2018b}, but cell-cell registration requires perturbations to the cell center that are parallel to the interface. In the SM we extend the 2D calculation to study the effects of parallel perturbations by changing the cell-cell registration. In the 2D toy model, we find that parallel perturbation incurs exactly the same energy penalty as a perpendicular perturbation up to linear order. Similar to the perpendicular perturbation, the parallel perturbation along the heterotypic interface requires the system to form new, high tension edges which are energetically very costly. Therefore, the cell centers are laterally pinned. In \ref{3Dcalc} we explore if de-registration in 3D systems creates a similar effect. We find that in an idealized hexagonal tissue, such a perturbation creates several new HIT facets, suggesting that a similar mechanism is operating in 3D.

{\it Mechanisms:} 
To better understand the pinning mechanism in 3D, we next develop two ordered toy models in 3D and study the response in the limit of zero fluctuations. Although the data shown so far, and work by some of us in 2D~\cite{Sussman2018b}, focused exclusively on the fluid-like regime with $s_0^{2D} > 3.81$ and $s_0^{3D} > 5.41$, we are also interested in how tissues close to the fluid-solid transition balance interfacial and bulk effects.  Therefore, our first toy model is initialized in a BCC lattice, as a ground state for the shape $s_0\sim5.31$. We fix $s_0=5.31$ to this ground state value, which we expect to be solid-like. This is similar to our recent work on ordered hexagonal monolayers~\cite{Sahu2020b}, but here we are investigating a different form of local perturbation in 3D.

\begin{figure}[htp]
\centering
\includegraphics[width=\columnwidth]{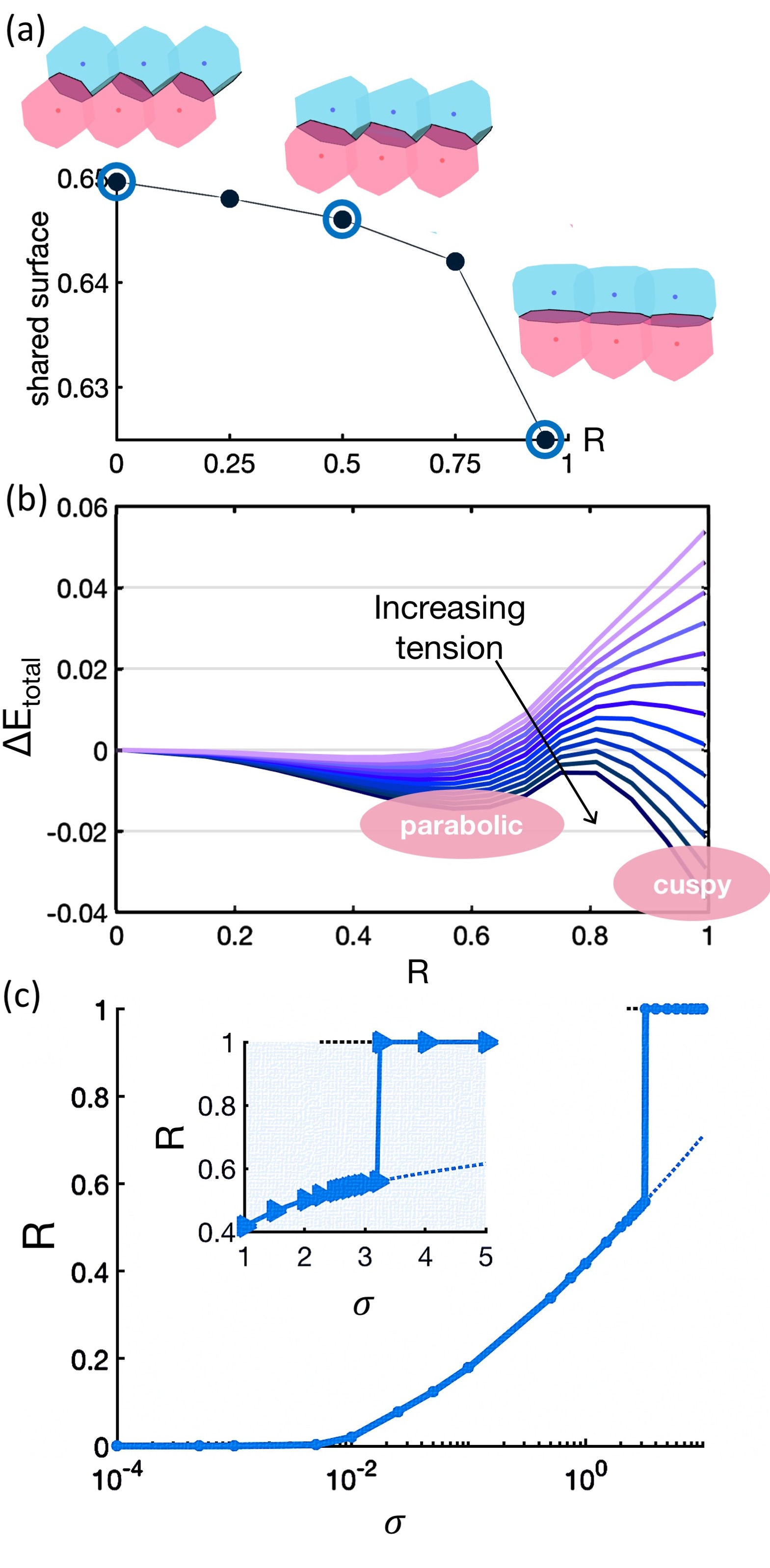}
\caption{\textbf{Global minimum becomes registered for higher tension}:(a) Shared surface area (shaded in dark for all snapshots) is computed as a function of the registration between the different cell types. The string of blue cells is allowed to move past the string of stationary pink cells. Snapshots for no, half and complete registration is shown for encircled points. (b) The change in the total energy of this system is plotted with respect to registry, for different values of tension ranging from 1 (pink) to 4 (blue) in increments of 0.25. (c) The ground-state registration is plotted for increasing tension. The solid curve represents the global minima. The dashed curves represent the local minima, parabolic in blue and cuspy in black. The inset zooms about the critical tension at which the transitions occurs between both types of minima. For all panels $s_0$ is set to $5.31$, the ground state for the BCC lattice.}
\label{fig:bcc_panel}
\end{figure}

In this work, we study the response of the system to a change in the registry $R$. In these ordered systems, we now use the lattice spacing as the lengthscale $l_0$ in the definition for registry in Eq.~\ref{eq:R}. We enable a string of polyhedra to slide past the string below (shown in Fig.~\ref{fig:bcc_panel}(a) insets), with an extra surface tension along the shared interfacial strip between both the sets of polyhedra. We compute the shared surface area and total energy of the system. We also find the global minima for different values of tension.  

We find that the shared surface area is minimized for increased registration values (shown in Fig.~\ref{fig:bcc_panel}(a)). While this suggests that perhaps complete registration is an energetically preferred state in some regimes, surprisingly, that is true only after a threshold value of tension. This can be seen by plotting the change in total energy with respect to increasing registration. One can observe two kinds of minima in the system \textemdash parabolic and `cuspy'~\cite{Sussman2018b}. While the former is common in particle-based models, with a spring-like potential locally around the minima, the later has a discontinuity in its slope due to topological pinning, which is expected from a 2D calculation as discussed in \ref{2Dcalc}. Physically, this means that the system experiences a steep linear rise in energy due to the formation of a new interfacial edge along the high tension cable, resulting in  discontinuous pinning forces that are independent of the magnitude of perturbation, as shown in Fig.~\ref{fig:bcc_extra}(b). The parabolic minima, on the other hand, have a continuous an linear restoring force, the slope proportional to the stiffness of the parabola. We plot the registration of energy minima as a function of interfacial tension in Fig.~\ref{fig:bcc_panel}(c).   

We observe two different branches corresponding to the two types of minima discussed before. For very low tension, shape preference dominates (as shown in Fig.~\ref{fig:bcc_extra2}(a)) and the system stays in BCC configuration. For moderate values of tension the system stays in the parabolic branch, but continuously transitions to non-zero registry. Just below the critical value of $\sigma_c\sim 3$, both the shape and interfacial energies become comparable (Fig.~\ref{fig:bcc_extra2}(b)) and at $\sigma_c$ the system discontinuously transitions to the tension-dominated- cuspy branch. The registration value also jumps to $R=1$. This branch originates at $\sigma\sim2$. 

Both kinds of minima become more stable as the tension increases. Stability is parameterized by the curvature or stiffness for parabolic minima in Fig.~\ref{fig:bcc_extra}(a) and by the linear slope or restoring force for the cuspy minima in Fig.~\ref{fig:bcc_extra}(c).  

In summary, for the solid-like toy model, we find that the physical mechanism that drives registration at high tension values is very similar to that of 2D systems. However, the story changes at lower values of tension where shape frustration begins to play a dominant role. This leads to minima that are partially-registered and exhibit a spring-like response to small perturbations, i.e. states are no longer topologically pinned. 

In 2D studies focused exclusively on fluid-like tissues~\cite{Sussman2018b}, only the cusp-like states were observed. This leads us to hypothesize that perhaps we observe a transition between normal, parabolic restoring forces to non-analytic cusp-y restoring forces in tissues that are more solid-like. Perhaps in more fluid-like systems the interfacial costs always dominate over the cost of cell shapes in the bulk, whereas in more solid-like systems the bulk effects dominate when the interfacial tension is low.

\begin{figure}[htp]
\centering
\includegraphics[width=\columnwidth]{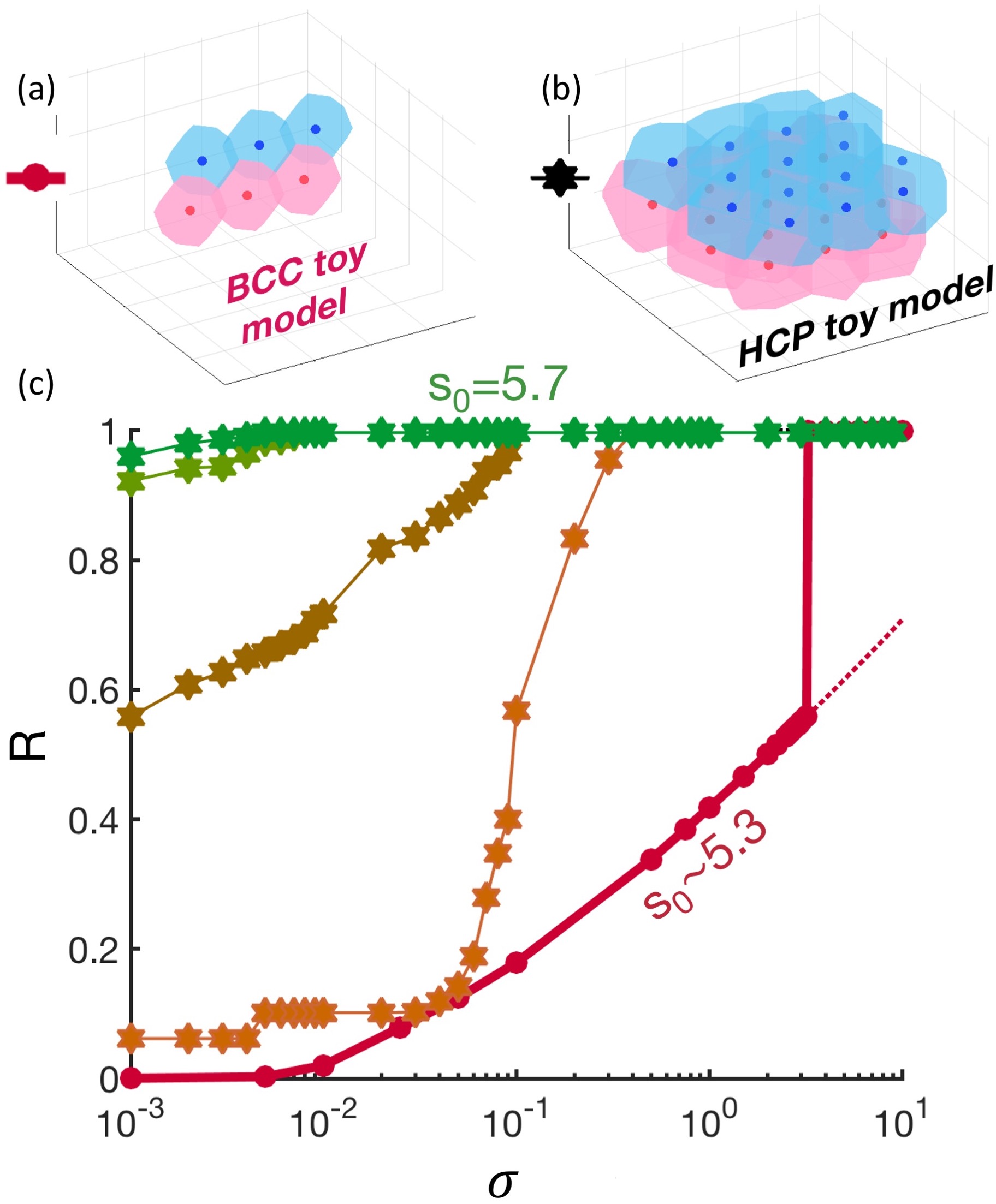}
\caption{\textbf{Shape frustration is less dominant for more fluid-like cell shapes}: (a,b) Snapshots for both the models are shown where both cell types are depicted in blue and red. Cell centers are depicted by solid spheres of respective cell color. (c) The ground-state registration for HCP toy model (filled hexagrams) is plotted with respect to tension, for increasing values of cell shape $s_0=5.4,5.5,5.6,5.7$ from orange to green. For comparison, the registration in BCC model (solid circles) is also shown here in red.  }
\label{fig:hcp_trial}
\end{figure}

To test this hypothesis, we develop a second toy model that is also ordered but not constrained to a string. Instead it is free to move along the 2D interface to change its registration. With this flexibility, we can explore the energetics of more elongated cell shapes like that of a uniform hexagonal prism ($s_0\sim5.72$). The interfacial cells can therefore be much more elonagated and fluid-like as compared to the minimal-perimeter-cells in a BCC lattice that have shape values as small as $s_0\sim5.31$. The interfacial layers are placed in HCP format as shown in Fig.~\ref{fig:hcp_trial}(b). 
There are buffer cells placed above and below the interface in a disordered fashion and are allowed to relax during the course of the perturbation.

Analogous to the the previous analysis, we compute the shared surface area and change in the energy profile with respect to registration, but this time for a wide range of cell shapes across the fluid-solid transition. We find that similar to the BCC toy model, the shared surface area decreases with registration as shown in Fig.~\ref{fig:hcp_extra}(a). For cell shapes near the rigidity transition, shape frustration plays a dominant role in the system's energy (Fig.~\ref{fig:hcp_extra}(b)). However, it becomes negligible for more fluid-like cell shapes as shown in Fig.~\ref{fig:hcp_trial}. This suggests that in fully disordered fluid-like systems, interfacial tension dominates shape preferences in determining interface geometry and response.

Finally, we verify the toy model predictions by first studying the geometry of disordered mechanically stable states in planar segregated HIT simulations, for different values of HIT and preferred cell shape. We let the system come to a steady state in the presence of fluctuations that can help the system find lower and lower energy metastable  minima in the complex potential energy landscape.

\begin{figure}[h!]
\centering
\includegraphics[width=\columnwidth]{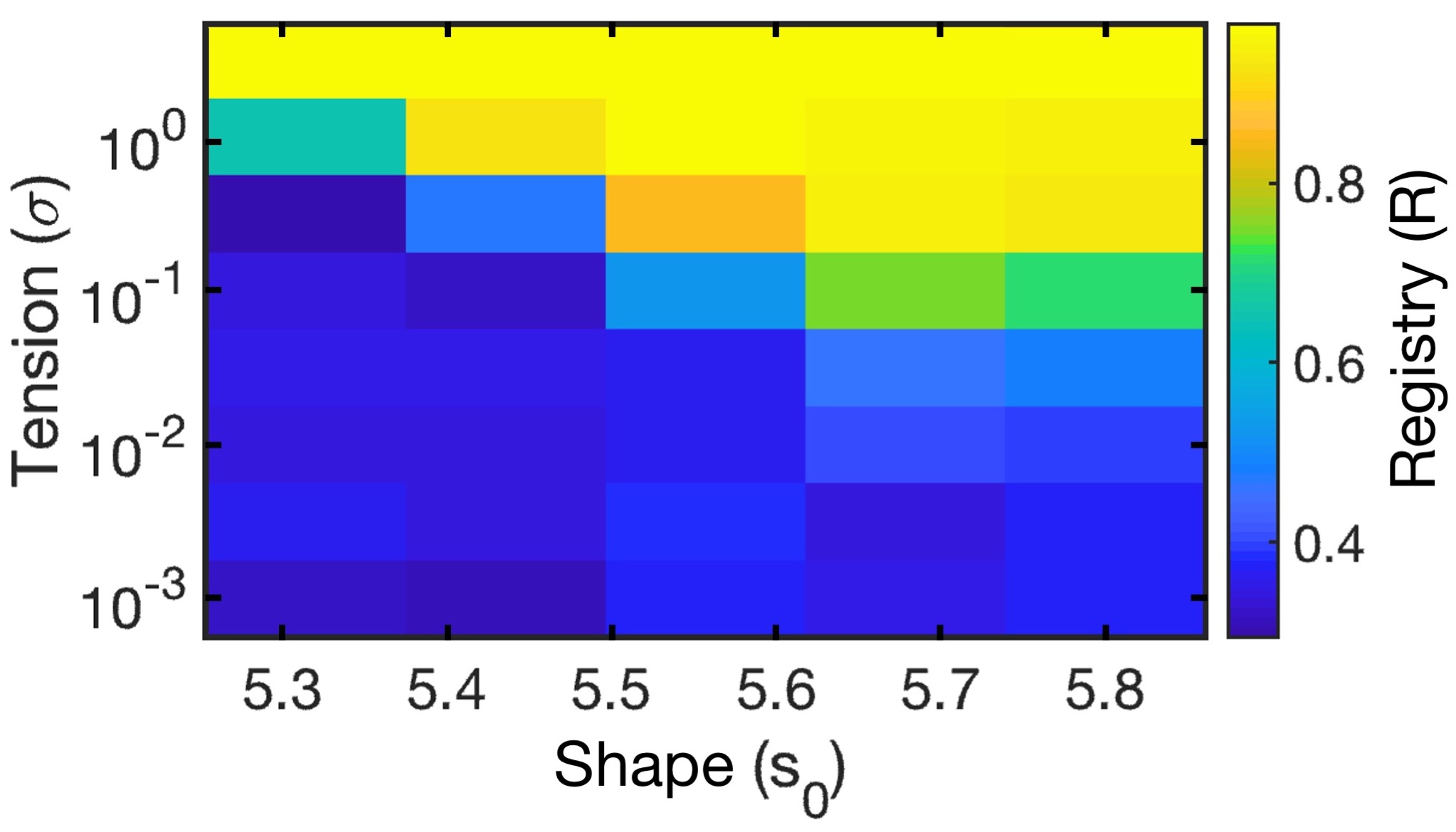}
\caption{\textbf{Transition to complete registration shifts for fluid-like cell shape in a disordered 3D simulation}: A heat-map for the average steady-state registration is shown as function of cell shape $s_0$ and interfacial tension $\sigma$. Yellow denotes complete registration and blue denotes partial registry. The registry between heterotypic cell-pairs is averaged over 200 different initializations. }
\label{fig:phaseDiag}
\end{figure}

Fig~\ref{fig:phaseDiag} shows the registration as a function of interfacial tension and target cell shape index $s_0$. The data demonstrate that just as in the toy models, the average registration rises rapidly to unity -- its maximum value -- when the interfacial tension increases above a threshold value. Moreover, for solid-like shapes the onset occurs at a higher threshold tension of order unity, while for fluid-like tissues the onset occurs when HIT is more than an order of magnitude lower than the typical tensions between homotypic cells.  In the supplement, we also confirm that the highly registered states are associated with cusp-like restoring forces, highlighted in Fig.~\ref{fig:SI_bulk}.

\section*{Discussion and Conclusions}
By studying a computational Voronoi model for confluent tissues in the presence of fluctuations, we show that three-dimensional binary mixtures, with heterotypic interfacial tension (HIT), sort robustly. This supports the claim that HIT is an important mechanism driving sharp compartmentalization during early embryonic development~\cite{Kesavan2020,Canty2017a}, and that it may be important for tissue segregation in other situations.  In addition to these collective dynamics, HIT also drives individual cells towards a prism-like geometry at the interface.  We find that the onset of these geometric signatures depends on a balance between the the magnitude of interfacial tension and constraints introduced by the the preferred cell shape that also govern the bulk tissue rheology. 

To understand the onset of these geometric signatures, we use cell-cell registration at the interface as a probe of the stability of these prism-like structures. We construct two simple toy models and study their energetics with respect to registry. In solid-like tissues, we find that for an interfacial tension $\sigma>\sigma_c$, the ground state is completely registered, which gives rise to a prismatic geometry. These states are topologically pinned, due to cusp-like pinning forces, previously observed in 2D mixtures with HIT. But for tensions $\sigma<\sigma_c$, interfacial energy is dominated by shape frustration and hence the ground state is less well-registered. The linear response of these minima is spring-like and not cuspy. Our data suggests that $\sigma_c$ decreases significantly as the tissue becomes more fluid-like.  This has important implications for development and tissue segregation, as it suggests that as tissues are tuned to be more solid-like, topological pinning at heterotypic interfaces is greatly reduced, thereby reducing the sharpness possible at compartment boundaries. In other words, it suggests the somewhat counterintuitive design principle for confluent systems that fluid-like rheologies lead to sharper interfaces.

We have also shown that a change in the magnitude of the interfacial tension can have a pronounced effect on the neighboring cellular geometry, by elongating interfacial cells into prism-like polyhedra oriented perpendicular to the interface. The observable facet areas also become larger. With the advancement of 3D segmentation techniques~\cite{Khan2014,Stegmaier2016,Browet2016,Fernandez-de-Manuel2017,Morales-Navarrete2019}, one can use these signatures as a toolkit to probe interfacial tensions in a 3D tissue, so that cells themselves can tell us about relative magnitudes of tissue surface tension nearby. 

One example case where this might be useful is in detecting the invasiveness of a carcinoma tumour. Our simple model would predict that if interfacial cell centers in the tumor are registered to those of the surrounding healthy tissue, then the interface has a higher surface tension and therefore it may be more unlikely for cells to exit the tumor and invade their surroundings. It would be interesting to see if there are any vestiges of this prediction that occur in pre-cancerous situations in real-world systems, such as Ductal carcinoma in situ (DCIS).

Another example is that of a stratified epithelium, where one can also study the interaction between two nearby tissue types, such as the basal and suprabasal layers, and look for geometric signatures across the interface. A prism-like geometry would strongly suggest the presence of an interfacial tension between these two tissue types. The prism-like geometry of cells can be visually detected using the En Face imaging technique\cite{Yokouchi2016,Rubsam2017}. As some current segmentation algorithms can also make predictions about 3D shape from random 2D cross-sections, these geometric signatures could potentially be characterized along the cross-sections~\cite{Sharp2019}. In general 3D tissue have complex interfacial geometries. Hence, one of the future avenues of this work would be to study the dependence on the curvature of the interface. 

Additional work should also focus on teasing apart how topological pinning affects dynamics in more general scenarios. For example, a natural extension of our general framework is to study two different cell shapes mixed together. After all, in realistic situations cells of two different tissue types likely also differ in preferred cell shape.  In 2D, some of us have determined that unique extrusion behaviour can emerge due to differential pinning of cells~\cite{Sahu2020a}, and something similar could occur in 3D. Topological pinning might be affecting the cell sorting dynamics as well. Presumably, less pinning can lead to seamless coarsening of nearby droplets, while more pinning can hinder the coalescence, and alter the sorting process. This would also be an interesting avenue for future study.

While our work demonstrates that changes to individual cellular geometries are a necessary consequence of HIT and tissue-scale segregation, these changes to cellular geometry could also be used as a signal to facilitate downstream patterning near interfaces. For example, the elongation of cells near a high-tension interface might trigger oriented cell divisions along the long-axis of these cells. We speculate that perhaps biology can make use of this subtle feedback for processes like targeted cellular proliferation during early phases of embryonic development.

\subsubsection*{Acknowledgement}

We thank Paula Sanematsu, Matthias Merkel, Daniel Sussman and Cristina Marchetti for helpful discussions, and M. Merkel for developing and sharing the original version of the 3D Voronoi code. This work was primarily funded by NSF-PoLS-1607416 and NSF-PoLS-2014192. PS and MLM acknowledge additional support from Simons Grant No. 454947.

\bibliographystyle{unsrtnat}
\bibliography{3Dtensionbib}

\begin{thebibliography}{67}
\providecommand{\natexlab}[1]{#1}
\providecommand{\url}[1]{\texttt{#1}}
\expandafter\ifx\csname urlstyle\endcsname\relax
  \providecommand{\doi}[1]{doi: #1}\else
  \providecommand{\doi}{doi: \begingroup \urlstyle{rm}\Url}\fi

\bibitem[Trinkaus and Groves(1955)]{Trinkaus1955a}
J.~P. Trinkaus and P.~W. Groves.
\newblock {Differentiation in culture of mixed aggregates of dissociated tissue
  cells}.
\newblock \emph{Proc. Natl. Acad. Sci.}, 1955.
\newblock ISSN 0027-8424.

\bibitem[Waites et~al.(2017)Waites, Cavaliere, Cachat, Danos, and
  Davies]{Waites2017b}
William Waites, Matteo Cavaliere, Elise Cachat, Vincent Danos, and Jamie~A.
  Davies.
\newblock {Organoid And Tissue Patterning Through Phase Separation: Use Of A
  Vertex Model To Relate Dynamics Of Patterning To Underlying Biophysical
  Parameters}.
\newblock \emph{bioRxiv}, 2017.

\bibitem[Unbekandt and Davies(2010)]{Unbekandt2010}
Mathieu Unbekandt and Jamie~A. Davies.
\newblock {Dissociation of embryonic kidneys followed by reaggregation allows
  the formation of renal tissues}.
\newblock \emph{Kidney Int.}, 2010.
\newblock ISSN 00852538.

\bibitem[Montero et~al.(2005)Montero, Carvalho, Wilsch-Br{\"{a}}uninger,
  Kilian, Mustafa, and Heisenberg]{Montero2005}
Juan~Antonio Montero, Lara Carvalho, Michaela Wilsch-Br{\"{a}}uninger, Beate
  Kilian, Chingdem Mustafa, and Carl~Philipp Heisenberg.
\newblock {Shield formation at the onset of zebrafish gastrulation}.
\newblock \emph{Development}, 2005.
\newblock ISSN 09501991.

\bibitem[Klopper et~al.(2010)Klopper, Krens, Grill, and
  Heisenberg]{Klopper2010}
A.~V. Klopper, G.~Krens, S.~W. Grill, and C.~P. Heisenberg.
\newblock {Finite-size corrections to scaling behavior in sorted cell
  aggregates}.
\newblock \emph{Eur. Phys. J. E}, 2010.
\newblock ISSN 12928941.

\bibitem[R{\"{u}}bsam et~al.(2017)R{\"{u}}bsam, Mertz, Kubo, Marg,
  J{\"{u}}ngst, Goranci-Buzhala, Schauss, Horsley, Dufresne, Moser, Ziegler,
  Amagai, Wickstr{\"{o}}m, and Niessen]{Rubsam2017}
Matthias R{\"{u}}bsam, Aaron~F. Mertz, Akiharu Kubo, Susanna Marg, Christian
  J{\"{u}}ngst, Gladiola Goranci-Buzhala, Astrid~C. Schauss, Valerie Horsley,
  Eric~R. Dufresne, Markus Moser, Wolfgang Ziegler, Masayuki Amagai, Sara~A.
  Wickstr{\"{o}}m, and Carien~M. Niessen.
\newblock {E-cadherin integrates mechanotransduction and EGFR signaling to
  control junctional tissue polarization and tight junction positioning}.
\newblock \emph{Nat. Commun.}, 2017.
\newblock ISSN 20411723.

\bibitem[Cochet-Escartin et~al.(2017)Cochet-Escartin, Locke, Shi, Steele, and
  Collins]{Cochet-Escartin2017a}
Olivier Cochet-Escartin, Tiffany~T. Locke, Winnie~H. Shi, Robert~E. Steele, and
  Eva Maria~S. Collins.
\newblock {Physical Mechanisms Driving Cell Sorting in Hydra}.
\newblock \emph{Biophys. J.}, 2017.
\newblock ISSN 15420086.

\bibitem[Friedl et~al.(2004)Friedl, Hegerfeldt, and Tusch]{Friedl2004}
Peter Friedl, Yael Hegerfeldt, and Miriam Tusch.
\newblock {Collective cell migration in morphogenesis and cancer}, 2004.
\newblock ISSN 02146282.

\bibitem[Foty and Steinberg(2005)]{Foty2005}
Ramsey~A. Foty and Malcolm~S. Steinberg.
\newblock {The differential adhesion hypothesis: A direct evaluation}.
\newblock \emph{Dev. Biol.}, 2005.
\newblock ISSN 00121606.

\bibitem[Pawlizak et~al.(2015)Pawlizak, Fritsch, Grosser, Ahrens, Thalheim,
  Riedel, Kie{\ss}ling, Oswald, Zink, Manning, and K{\"{a}}s]{Pawlizak2015}
Steve Pawlizak, Anatol~W. Fritsch, Steffen Grosser, Dave Ahrens, Tobias
  Thalheim, Stefanie Riedel, Tobias~R. Kie{\ss}ling, Linda Oswald, Mareike
  Zink, M.~Lisa Manning, and Josef~A. K{\"{a}}s.
\newblock {Testing the differential adhesion hypothesis across the
  epithelial-mesenchymal transition}.
\newblock \emph{New J. Phys.}, 17:\penalty0 091001, 2015.
\newblock ISSN 13672630.

\bibitem[Song et~al.(2016)Song, Tung, Lu, Pardo, Wu, Das, Kao, Chen, and
  Ma]{Song2016}
Wei Song, Chih~Kuan Tung, Yen~Chun Lu, Yehudah Pardo, Mingming Wu, Moumita Das,
  Der~I. Kao, Shuibing Chen, and Minglin Ma.
\newblock {Dynamic self-organization of microwell-aggregated cellular
  mixtures}.
\newblock \emph{Soft Matt.}, 12:\penalty0 5739, 2016.
\newblock ISSN 17446848.

\bibitem[Turing(1990)]{Turing1990}
A.~M. Turing.
\newblock {The chemical basis of morphogenesis}.
\newblock \emph{Bull. Math. Biol.}, 1990.
\newblock ISSN 00928240.

\bibitem[Streichan et~al.(2018)Streichan, Lefebvre, Noll, Wieschaus, and
  Shraiman]{Streichan2018}
Sebastian~J Streichan, Matthew~F Lefebvre, Nicholas Noll, Eric~F Wieschaus, and
  Boris~I Shraiman.
\newblock {Global morphogenetic flow is accurately predicted by the spatial
  distribution of myosin motors}.
\newblock \emph{Elife}, 2018.

\bibitem[Steinberg(1963)]{Steinberg1963a}
Malcolm~S. Steinberg.
\newblock {Reconstruction of tissues by dissociated cells}.
\newblock \emph{Sci. Sci.}, 1963.
\newblock ISSN 00368075.

\bibitem[Steinberg(2007)]{STEINBERG2007281}
Malcolm~S Steinberg.
\newblock Differential adhesion in morphogenesis: a modern view.
\newblock \emph{Current Opinion in Genetics and Development}, 17\penalty0
  (4):\penalty0 281 -- 286, 2007.
\newblock ISSN 0959-437X.
\newblock URL
  \url{http://www.sciencedirect.com/science/article/pii/S0959437X07001062}.
\newblock Pattern formation and developmental mechanisms.

\bibitem[Harris(1976)]{Harris1976b}
Albert~K. Harris.
\newblock {Is cell sorting caused by differences in the work of intercellular
  adhesion? A critique of the steinberg hypothesis}.
\newblock \emph{J. Theor. Biol.}, 1976.
\newblock ISSN 10958541.

\bibitem[Brodland(2002)]{Brodland2002a}
G.~Wayne Brodland.
\newblock {The Differential Interfacial Tension Hypothesis (DITH): A
  comprehensive theory for the self-rearrangement of embryonic cells and
  tissues}.
\newblock \emph{J. Biomech. Eng.}, 2002.
\newblock ISSN 01480731.

\bibitem[Mertz et~al.(2012)Mertz, Banerjee, Che, German, Xu, Hyland, Marchetti,
  Horsley, and Dufresne]{Mertz2012a}
Aaron~F. Mertz, Shiladitya Banerjee, Yonglu Che, Guy~K. German, Ye~Xu, Callen
  Hyland, M.~Cristina Marchetti, Valerie Horsley, and Eric~R. Dufresne.
\newblock {Scaling of traction forces with the size of cohesive cell colonies}.
\newblock \emph{Phys. Rev. Lett.}, 2012.
\newblock ISSN 00319007.

\bibitem[Ma{\^{i}}tre et~al.(2012)Ma{\^{i}}tre, Berthoumieux, Krens, Salbreux,
  J{\"{u}}licher, Paluch, and Heisenberg]{Maitre2012b}
Jean~L{\'{e}}on Ma{\^{i}}tre, H{\'{e}}l{\`{e}}ne Berthoumieux, Simon
  Frederik~Gabriel Krens, Guillaume Salbreux, Frank J{\"{u}}licher, Ewa Paluch,
  and Carl~Philipp Heisenberg.
\newblock {Adhesion functions in cell sorting by mechanically coupling the
  cortices of adhering cells}.
\newblock \emph{Science (80-. ).}, 2012.
\newblock ISSN 10959203.

\bibitem[Amack and Manning(2012)]{Amack2012a}
Jeffrey~D. Amack and M.~Lisa Manning.
\newblock {Knowing the boundaries: Extending the differential adhesion
  hypothesis in embryonic cell sorting}, 2012.
\newblock ISSN 10959203.

\bibitem[Manning et~al.(2010{\natexlab{a}})Manning, Foty, Steinberg, and
  Schoetz]{Manning2010c}
M.~Lisa Manning, Ramsey~A. Foty, Malcolm~S. Steinberg, and Eva~Maria Schoetz.
\newblock {Coaction of intercellular adhesion and cortical tension specifies
  tissue surface tension}.
\newblock \emph{Proc. Natl. Acad. Sci. U. S. A.}, 2010{\natexlab{a}}.
\newblock ISSN 00278424.

\bibitem[Engl et~al.(2014)Engl, Arasi, Yap, Thiery, and Viasnoff]{Engl2014}
W.~Engl, B.~Arasi, L.~L. Yap, J.~P. Thiery, and V.~Viasnoff.
\newblock {Actin dynamics modulate mechanosensitive immobilization of
  E-cadherin at adherens junctions}.
\newblock \emph{Nat. Cell Biol.}, 2014.
\newblock ISSN 14764679.

\bibitem[Graner and Glazier(1992)]{Graner1992}
Fran{\c{c}}ois Graner and James~A. Glazier.
\newblock {Simulation of biological cell sorting using a two-dimensional
  extended Potts model}.
\newblock \emph{Phys. Rev. Lett.}, 69:\penalty0 213--216, 1992.
\newblock ISSN 00319007.

\bibitem[Barton et~al.(2017)Barton, Henkes, Weijer, and R.]{Barton2017}
D.~L. Barton, S.~Henkes, C.~J. Weijer, and Sknepnek R.
\newblock Active vertex model for cell-resolution description of epithelial
  tissue mechanics.
\newblock \emph{PLoS Comput. Biol.}, 13:\penalty0 e1005569, 2017.

\bibitem[Canty et~al.(2017)Canty, Zarour, Kashkooli, Fran{\c{c}}ois, and
  Fagotto]{Canty2017a}
Laura Canty, Eleyine Zarour, Leily Kashkooli, Paul Fran{\c{c}}ois, and
  Fran{\c{c}}ois Fagotto.
\newblock {Sorting at embryonic boundaries requires high heterotypic
  interfacial tension}.
\newblock \emph{Nat. Comm.}, 8:\penalty0 157, 2017.

\bibitem[Sussman et~al.(2018)Sussman, Schwarz, Marchetti, and
  Manning]{Sussman2018b}
Daniel~M. Sussman, J.~M. Schwarz, M.~Cristina Marchetti, and M.~Lisa Manning.
\newblock {Soft yet Sharp Interfaces in a Vertex Model of Confluent Tissue}.
\newblock \emph{Phys. Rev. Lett.}, 120:\penalty0 058001, 2018.

\bibitem[Flenner et~al.(2012)Flenner, Janosi, Barz, Neagu, Forgacs, and
  Kosztin]{PhysRevE.85.031907}
Elijah Flenner, Lorant Janosi, Bogdan Barz, Adrian Neagu, Gabor Forgacs, and
  Ioan Kosztin.
\newblock Kinetic monte carlo and cellular particle dynamics simulations of
  multicellular systems.
\newblock \emph{Phys. Rev. E}, 85:\penalty0 031907, Mar 2012.

\bibitem[Merks and Glazier(2005)]{Merks2005}
Roeland~M.H. Merks and James~A. Glazier.
\newblock {A cell-centered approach to developmental biology}.
\newblock \emph{Phys. A Stat. Mech. its Appl.}, 2005.
\newblock ISSN 03784371.

\bibitem[Sun and Wang(2013)]{Sun2013}
Yi~Sun and Qi~Wang.
\newblock {Modeling and simulations of multicellular aggregate self-assembly in
  biofabrication using kinetic Monte Carlo methods}.
\newblock \emph{Soft Matter}, 2013.
\newblock ISSN 17446848.

\bibitem[Jiang et~al.(1998)Jiang, Levine, and Glazier]{Jiang1998}
Yi~Jiang, Herbert Levine, and James Glazier.
\newblock {Possible cooperation of differential adhesion and chemotaxis in
  mound formation of Dictyostelium}.
\newblock \emph{Biophys. J.}, 1998.
\newblock ISSN 00063495.

\bibitem[Palsson(2008)]{Palsson2008}
Eirikur Palsson.
\newblock {A 3-D model used to explore how cell adhesion and stiffness affect
  cell sorting and movement in multicellular systems}.
\newblock \emph{J. Theor. Biol.}, 2008.
\newblock ISSN 00225193.

\bibitem[Landsberg et~al.(2009)Landsberg, Farhadifar, Ranft, Umetsu, Widmann,
  Bittig, Said, J{\"{u}}licher, and Dahmann]{Landsberg2009a}
Katharina~P. Landsberg, Reza Farhadifar, Jonas Ranft, Daiki Umetsu, Thomas~J.
  Widmann, Thomas Bittig, Amani Said, Frank J{\"{u}}licher, and Christian
  Dahmann.
\newblock {Increased Cell Bond Tension Governs Cell Sorting at the Drosophila
  Anteroposterior Compartment Boundary}.
\newblock \emph{Curr. Biol.}, 2009.
\newblock ISSN 09609822.

\bibitem[Manning et~al.(2010{\natexlab{b}})Manning, Foty, Steinberg, and
  Schoetz]{Manning2010b}
M.~Lisa Manning, Ramsey~A. Foty, Malcolm~S. Steinberg, and Eva-Maria Schoetz.
\newblock {Coaction of intercellular adhesion and cortical tension specifies
  tissue surface tension}.
\newblock \emph{Proc. Natl. Acad. Sci.}, 2010{\natexlab{b}}.
\newblock ISSN 0027-8424.

\bibitem[Dahmann et~al.(2011)Dahmann, Oates, and Brand]{Dahmann2011}
Christian Dahmann, Andrew~C. Oates, and Michael Brand.
\newblock {Boundary formation and maintenance in tissue development}, 2011.
\newblock ISSN 14710056.

\bibitem[Nnetu et~al.(2012)Nnetu, Knorr, K{\"{a}}s, and Zink]{Nnetu2012a}
Kenechukwu~David Nnetu, Melanie Knorr, Josef K{\"{a}}s, and Mareike Zink.
\newblock {The impact of jamming on boundaries of collectively moving
  weak-interacting cells}.
\newblock \emph{New J. Phys.}, 2012.
\newblock ISSN 13672630.

\bibitem[Monier et~al.(2011)Monier, P{\'{e}}lissier-Monier, and
  Sanson]{Monier2011}
Bruno Monier, Anne P{\'{e}}lissier-Monier, and B{\'{e}}n{\'{e}}dicte Sanson.
\newblock {Establishment and maintenance of compartmental boundaries: Role of
  contractile actomyosin barriers}, 2011.
\newblock ISSN 1420682X.

\bibitem[Calzolari et~al.(2014)Calzolari, Terriente, and
  Pujades]{Calzolari2014}
Simone Calzolari, Javier Terriente, and Cristina Pujades.
\newblock {Cell segregation in the vertebrate hindbrain relies on actomyosin
  cables located at the interhombomeric boundaries}.
\newblock \emph{EMBO J.}, 2014.
\newblock ISSN 14602075.

\bibitem[Kesavan et~al.(2020)Kesavan, Machate, Hans, and Brand]{Kesavan2020}
Gokul Kesavan, Anja Machate, Stefan Hans, and Michael Brand.
\newblock {Cell-fate plasticity, adhesion and cell sorting complementarily
  establish a sharp midbrain-hindbrain boundary}.
\newblock \emph{Development}, 2020.
\newblock ISSN 14779129.

\bibitem[Sahu et~al.(2020{\natexlab{a}})Sahu, Kang, Erdemci-Tandogan, and
  Manning]{Sahu2020b}
Preeti Sahu, Janice Kang, Gonca Erdemci-Tandogan, and M.~Lisa Manning.
\newblock {Linear and nonlinear mechanical responses can be quite different in
  models for biological tissues}.
\newblock \emph{Soft Matter}, 2020{\natexlab{a}}.
\newblock ISSN 17446848.

\bibitem[Hutson et~al.(2008)Hutson, Brodland, Yang, and
  Viens]{PhysRevLett.101.148105}
M.~Shane Hutson, G.~Wayne Brodland, Justina Yang, and Denis Viens.
\newblock Cell sorting in three dimensions: Topology, fluctuations, and
  fluidlike instabilities.
\newblock \emph{Phys. Rev. Lett.}, 101:\penalty0 148105, Oct 2008.

\bibitem[Sugimura et~al.(2016)Sugimura, Lenne, and Graner]{Sugimura2016}
Kaoru Sugimura, Pierre~Fran{\c{c}}ois Lenne, and Fran{\c{c}}ois Graner.
\newblock {Measuring forces and stresses in situ in living tissues}.
\newblock \emph{Dev.}, 2016.
\newblock ISSN 14779129.

\bibitem[Tsai et~al.(2019)Tsai, Sikora, Xia, Colak-Champollion, Knaut,
  Heisenberg, and Megason]{Tsai2019}
Tony Y.-C. Tsai, Mateusz Sikora, Peng Xia, Tugba Colak-Champollion, Holger
  Knaut, Carl-Philipp Heisenberg, and Sean~G. Megason.
\newblock {An adhesion code ensures robust pattern formation during tissue
  morphogenesis}.
\newblock \emph{bioRxiv}, 2019.

\bibitem[Okuda et~al.(2015)Okuda, Inoue, and Adachi]{Okuda2015}
Satoru Okuda, Yasuhiro Inoue, and Taiji Adachi.
\newblock {Three-dimensional vertex model for simulating multicellular
  morphogenesis}.
\newblock \emph{Biophys. Physicobiology}, 2015.
\newblock ISSN 2189-4779.

\bibitem[Yang and Brodland(2009)]{Yang2009}
Justina Yang and G.~Wayne Brodland.
\newblock {Estimating interfacial tension from the shape histories of cells in
  compressed aggregates: A computational study}.
\newblock \emph{Ann. Biomed. Eng.}, 2009.
\newblock ISSN 00906964.

\bibitem[{Canongia Lopes}(2002)]{CanongiaLopes2002}
Jos{\'{e}}~N. {Canongia Lopes}.
\newblock {Microphase separation in mixtures of Lennard-Jones particles}.
\newblock In \emph{Phys. Chem. Chem. Phys.}, 2002.

\bibitem[Majumder and Das(2011)]{Majumder2011}
Suman Majumder and Subir~K. Das.
\newblock {Universality in fluid domain coarsening: The case of vapor-liquid
  transition}.
\newblock \emph{EPL}, 2011.
\newblock ISSN 02955075.

\bibitem[Chremos et~al.(2014)Chremos, Nikoubashman, and
  Panagiotopoulos]{Chremos2014}
Alexandros Chremos, Arash Nikoubashman, and Athanassios~Z. Panagiotopoulos.
\newblock {Flory-Huggins parameter $\chi$, from binary mixtures of
  Lennard-Jones particles to block copolymer melts}.
\newblock \emph{J. Chem. Phys.}, 2014.
\newblock ISSN 00219606.

\bibitem[Nagai and Honda(2001)]{Nagai2001}
Tatsuzo Nagai and Hisao Honda.
\newblock {A dynamic cell model for the formation of epithelial tissues}.
\newblock \emph{Phil. Mag. B: Phys. Cond. Matt., Stat. Mech., Elec. Opt. Mag.
  Prop.}, 81:\penalty0 699--719, 2001.

\bibitem[Farhadifar et~al.(2007)Farhadifar, R{\"{o}}per, Aigouy, Eaton, and
  J{\"{u}}licher]{Farhadifar2007a}
Reza Farhadifar, Jens~Christian R{\"{o}}per, Benoit Aigouy, Suzanne Eaton, and
  Frank J{\"{u}}licher.
\newblock {The Influence of Cell Mechanics, Cell-Cell Interactions, and
  Proliferation on Epithelial Packing}.
\newblock \emph{Curr. Biol.}, 17\penalty0 (24):\penalty0 2095--2104, 2007.
\newblock ISSN 09609822.

\bibitem[Teleman et~al.(2007)Teleman, Hufnagel, Rouault, Shraiman, and
  Cohen]{Teleman2007}
A.~A. Teleman, L.~Hufnagel, H.~Rouault, B.~I. Shraiman, and S.~M. Cohen.
\newblock {On the mechanism of wing size determination in fly development}.
\newblock \emph{Proc. Natl. Acad. Sci.}, 2007.
\newblock ISSN 0027-8424.

\bibitem[Staple et~al.(2010)Staple, Farhadifar, R{\"{o}}per, Aigouy, Eaton, and
  J{\"{u}}licher]{Staple2010a}
D.~B. Staple, R.~Farhadifar, J.~C. R{\"{o}}per, B.~Aigouy, S.~Eaton, and
  F.~J{\"{u}}licher.
\newblock {Mechanics and remodelling of cell packings in epithelia}.
\newblock \emph{Eur. Phys. J. E}, 33:\penalty0 117--127, 2010.

\bibitem[Hilgenfeldt et~al.(2008)Hilgenfeldt, Erisken, and
  Carthew]{Hilgenfeldt2008}
S.~Hilgenfeldt, S.~Erisken, and R.~W. Carthew.
\newblock {Physical modeling of cell geometric order in an epithelial tissue}.
\newblock \emph{Proc. Natl. Acad. Sci.}, 2008.
\newblock ISSN 0027-8424.

\bibitem[Chiou et~al.(2012)Chiou, Hufnagel, and Shraiman]{Chiou2012}
Kevin~K. Chiou, Lars Hufnagel, and Boris~I. Shraiman.
\newblock {Mechanical stress inference for two dimensional cell arrays}.
\newblock \emph{PLoS Comput. Biol.}, 2012.
\newblock ISSN 1553734X.

\bibitem[Bi et~al.(2015)Bi, Lopez, Schwarz, and Manning]{Bi2015c}
Dapeng Bi, J.~H. Lopez, J.~M. Schwarz, and M.~Lisa Manning.
\newblock {A density-independent rigidity transition in biological tissues}.
\newblock \emph{Nat. Phys.}, 11\penalty0 (12):\penalty0 1074--1079, 2015.
\newblock ISSN 17452481.

\bibitem[Fletcher et~al.(2014)Fletcher, Osterfield, Baker, and
  Shvartsman]{Fletcher2014}
Alexander~G. Fletcher, Miriam Osterfield, Ruth~E. Baker, and Stanislav~Y.
  Shvartsman.
\newblock {Vertex models of epithelial morphogenesis}, 2014.
\newblock ISSN 15420086.

\bibitem[Merkel and Manning(2018)]{Merkel2018}
Matthias Merkel and M.~Lisa Manning.
\newblock {A geometrically controlled rigidity transition in a model for
  confluent 3D tissues}.
\newblock \emph{New J. Phys.}, 2018.
\newblock ISSN 13672630.

\bibitem[Bi et~al.(2016)Bi, Yang, Marchetti, and Manning]{Bi2016b}
Dapeng Bi, Xingbo Yang, M.~Cristina Marchetti, and M.~Lisa Manning.
\newblock {Motility-driven glass and jamming transitions in biological
  tissues}.
\newblock \emph{Phys. Rev. X}, 2016.
\newblock ISSN 21603308.

\bibitem[Lucarini(2009)]{Lucarini2009}
Valerio Lucarini.
\newblock {Three-dimensional random voronoi tessellations: From cubic crystal
  lattices to poisson point processes}.
\newblock \emph{J. Stat. Phys.}, 2009.
\newblock ISSN 00224715.

\bibitem[Sahu et~al.(2020{\natexlab{b}})Sahu, Sussman, R{\"{u}}bsam, Mertz,
  Horsley, Dufresne, Niessen, Marchetti, Manning, and Schwarz]{Sahu2020a}
Preeti Sahu, Daniel~M. Sussman, Matthias R{\"{u}}bsam, Aaron~F. Mertz, Valerie
  Horsley, Eric~R. Dufresne, Carien~M. Niessen, M.~Cristina Marchetti, M.~Lisa
  Manning, and J.~M. Schwarz.
\newblock {Small-scale demixing in confluent biological tissues}.
\newblock \emph{Soft Matter}, 2020{\natexlab{b}}.
\newblock ISSN 17446848.

\bibitem[Khan et~al.(2014)Khan, Wang, Wieschaus, and Kaschube]{Khan2014}
Zia Khan, Yu~Chiun Wang, Eric~F. Wieschaus, and Matthias Kaschube.
\newblock {Quantitative 4D analyses of epithelial folding during Drosophila
  gastrulation}.
\newblock \emph{Dev.}, 2014.
\newblock ISSN 14779129.

\bibitem[Stegmaier et~al.(2016)Stegmaier, Amat, Lemon, McDole, Wan, Teodoro,
  Mikut, and Keller]{Stegmaier2016}
Johannes Stegmaier, Fernando Amat, William~C. Lemon, Katie McDole, Yinan Wan,
  George Teodoro, Ralf Mikut, and Philipp~J. Keller.
\newblock {Real-Time Three-Dimensional Cell Segmentation in Large-Scale
  Microscopy Data of Developing Embryos}.
\newblock \emph{Dev. Cell}, 2016.
\newblock ISSN 18781551.

\bibitem[Browet et~al.(2016)Browet, {De Vleeschouwer}, Jacques, Mathiah,
  Saykali, and Migeotte]{Browet2016}
A.~Browet, C.~{De Vleeschouwer}, L.~Jacques, N.~Mathiah, B.~Saykali, and
  I.~Migeotte.
\newblock {Cell segmentation with random ferns and graph-cuts}.
\newblock In \emph{Proc. - Int. Conf. Image Process. ICIP}, 2016.
\newblock ISBN 9781467399616.

\bibitem[Fern{\'{a}}ndez-de Man{\'{u}}el et~al.(2017)Fern{\'{a}}ndez-de
  Man{\'{u}}el, D{\'{i}}az-D{\'{i}}az, Jim{\'{e}}nez-Carretero, Torres, and
  Montoya]{Fernandez-de-Manuel2017}
Laura Fern{\'{a}}ndez-de Man{\'{u}}el, Covadonga D{\'{i}}az-D{\'{i}}az, Daniel
  Jim{\'{e}}nez-Carretero, Miguel Torres, and Mar{\'{i}}a~C. Montoya.
\newblock {ESC-track: A computer workflow for 4-D segmentation, tracking,
  lineage tracing and dynamic context analysis of ESCs}.
\newblock \emph{Biotechniques}, 2017.
\newblock ISSN 19409818.

\bibitem[Morales-Navarrete et~al.(2019)Morales-Navarrete, Nonaka, Scholich,
  Segovia-Miranda, de~Back, Meyer, Bogorad, Koteliansky, Brusch, Kalaidzidis,
  J{\"{u}}licher, Friedrich, and Zerial]{Morales-Navarrete2019}
Hern{\'{a}}n Morales-Navarrete, Hidenori Nonaka, Andr{\'{e}} Scholich,
  Fabi{\'{a}}n Segovia-Miranda, Walter de~Back, Kirstin Meyer, Roman~L.
  Bogorad, Victor Koteliansky, Lutz Brusch, Yannis Kalaidzidis, Frank
  J{\"{u}}licher, Benjamin~M. Friedrich, and Marino Zerial.
\newblock {Liquid-crystal organization of liver tissue}.
\newblock \emph{Elife}, 2019.
\newblock ISSN 2050084X.

\bibitem[Yokouchi et~al.(2016)Yokouchi, Atsugi, {Van Logtestijn}, Tanaka,
  Kajimura, Suematsu, Furuse, Amagai, and Kubo]{Yokouchi2016}
Mariko Yokouchi, Toru Atsugi, Mark {Van Logtestijn}, Reiko~J. Tanaka, Mayumi
  Kajimura, Makoto Suematsu, Mikio Furuse, Masayuki Amagai, and Akiharu Kubo.
\newblock {Epidermal cell turnover across tight junctions based on Kelvin's
  tetrakaidecahedron cell shape}.
\newblock \emph{Elife}, 2016.
\newblock ISSN 2050084X.

\bibitem[Sharp et~al.(2019)Sharp, Merkel, {Lisa Manning}, and Liu]{Sharp2019}
Tristan~A. Sharp, Matthias Merkel, M.~{Lisa Manning}, and Andrea~J. Liu.
\newblock {Inferring statistical properties of 3D cell geometry from 2D
  slices}.
\newblock \emph{PLoS One}, 2019.
\newblock ISSN 19326203.

\bibitem[Marchetti et~al.(2013)Marchetti, Joanny, Ramaswamy, Liverpool, Prost,
  Rao, and Simha]{RevModPhys.85.1143}
M.~C. Marchetti, J.~F. Joanny, S.~Ramaswamy, T.~B. Liverpool, J.~Prost, Madan
  Rao, and R.~Aditi Simha.
\newblock Hydrodynamics of soft active matter.
\newblock \emph{Rev. Mod. Phys.}, 85:\penalty0 1143--1189, Jul 2013.
\newblock URL \url{https://link.aps.org/doi/10.1103/RevModPhys.85.1143}.

\end{thebibliography}


\pagebreak
\section*{Electronic Supplemental Material}
\beginsupplement

\subsection{\label{3dspp}Details about self-propulsion dynamics}
For dynamical simulations, we use the static model developed for 3D confluent tissues~\cite{Merkel2018}, but with added cellular activity. The model essentially allows active fluctuations during time evolution of cells. Cells are polyhedra derived from a space-filling Voronoi tessellation of the periodic simulation box. The energy functional given by Eq.~\ref{eq:HST_3}, provides the mechanical force $\mathbf{f}_i=\partial e/\partial \mathbf{r}_i$ on cells due to changes in cell shape and/or shared interface area. For introducing activity in this dynamics, one needs to choose the frame of reference --  for example a static extra-cellular fluid. In a self-propelled system, the $i_{th}$ cell has a polarization vector $\hat{\mathbf{n}}_i$ and exerts an active force of $v_0/\mu$ on the static media, where $\mu$ is the mobility of the cell and $v_0$ is the self-propulsion speed. Setting $\mu=1$, the dynamical equation is :
\begin{equation}
    \label{eq:spp3}
    \dfrac{d\mathbf{r}_i}{dt}=v_0 \hat{\mathbf{n}}_i + \mathbf{f}_i.
\end{equation}
The polarization vector evolves via white Gaussian noise on a unit sphere with diffusion coefficient $D_r$, or 
\begin{equation}
    \label{eq:RotDif}
    \dfrac{d\hat{\mathbf{n}}_i}{dt}=\sqrt{2D_r}\big(\mathbf{E}-\hat{\mathbf{n}}_i\hat{\mathbf{n}}_i\big)\dot \xi_i,
\end{equation}
where $\mathbf{E}$ is the $3\times 3$ identity tensor, the dyadic product of the polarization vector with itself is given by $\hat{\mathbf{n}}_i\hat{\mathbf{n}}_i$ and $\xi_i$ is the white Gaussian noise with $\langle \xi_i\rangle=0$ and $\langle \xi_i\rangle(t)\langle \xi_j\rangle(t')=\delta_{ij}\delta(t-t')\mathbf{E}$.

Timescales in our dynamic equations are nondimensionalized by the natural time unit $\Tilde{t}=1/K_{V}V_{0}^{4/3}$, which is unity for our choice of parameter values. We use a self-propulsion speed of $v_0=0.1$ and a rotational diffusion coefficient $D_r=1.0 $. For such high values of rotational diffusion, the transition to a Brownian regime happens rather quickly, i.e $t/\Tilde{t}>1/D_r$.  In the absence of any cell-cell interactions, a self-propelled particle with that choice of parameters would diffuse over a characteristic self-diffusivity timescale $100\Tilde{t}$~\cite{RevModPhys.85.1143}. However, cell-cell interactions slow down diffusion, and we find that in our model the characteristic diffusion timescale for a fixed shape index of 5.5 is $\tau_s^0 \equiv 1000\Tilde{t}$. Therefore, for simulations in the presence of fluctuations, we use an integration step size $\Delta t=0.01 \Tilde{t}$ and a time-span $T_{SS}$ of $1 \times 10^5$ steps which is sufficient to allow a system of this size to achieve a steady state. While this holds true for the entire range of the explored tension, values in the lower-moderate regime ($\sigma<1$), achieve a steady state sooner, i.e. in just $1 \times 10^4$ 10000 steps. In the sorting simulations, we extend this timescale even further to allow cells to travel across the length of the simulation box more than 60 times, ensuring that even long-timescale patterning processes are allowed. 
\subsection{Maximum demixing value with respect to system-size}
For a small system-size, the demixing cannot attain the maximum possible value of unity as the number of heterotypic facets is not negligible as compared to the number of homotypic facets. Therefore we look at demixing in a segregated arrangement as a function of system-size.

\begin{figure}[htp]
\centering
\includegraphics[width=\columnwidth]{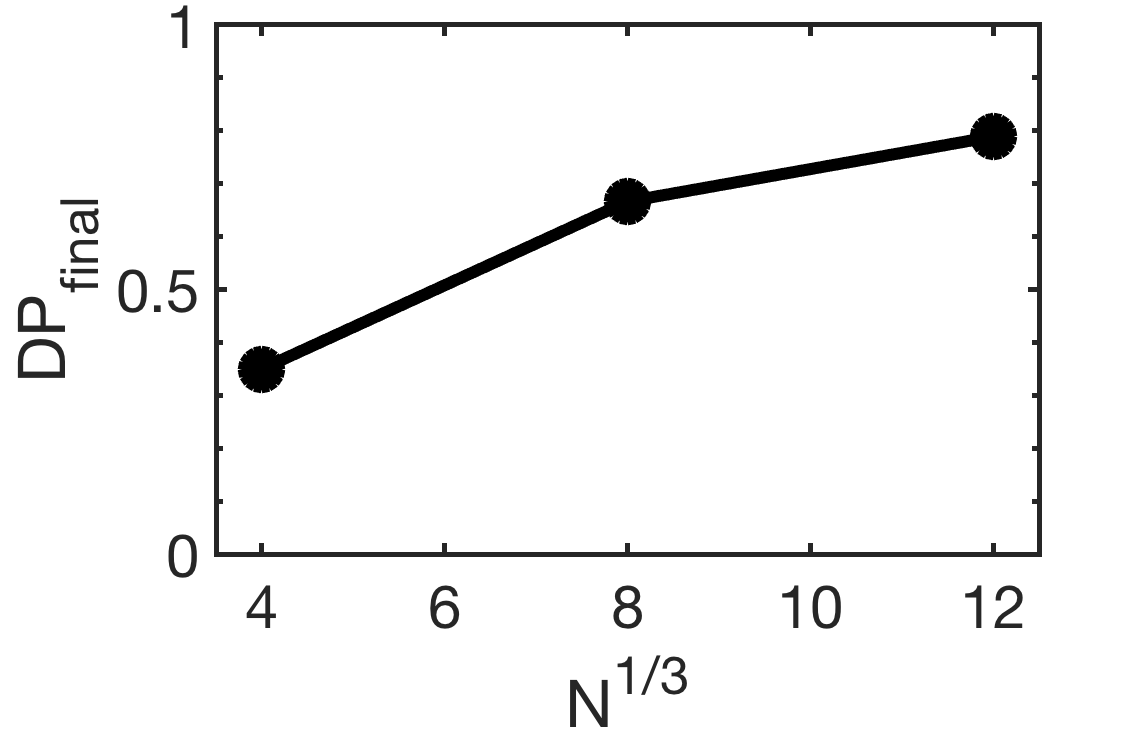}
\caption{Plot for the observed final demixing parameter in simulations ($DP_{final}$) as a function of the cubic root of the system size. Data is consistent with a straight line. 
}
\label{fig:dp_max}
\end{figure}

The simple assumption that a given interfacial cell shares only one-third of its facets with heterotypic neighbors, i.e. $DP_{final} \propto N^{1/3}$ is in fairly good agreement with the observed final demixing values seen in our simulations as shown in Fig~\ref{fig:dp_max}. 

The small deviations from this prediction we see in simulations are not surprising; even though the bilayer geometry is the least-surface-area configuration for a cubic simulation space like ours, for a mixed initial conditions with periodic boundary conditions it is actually difficult to achieve this configuration. In general, compartments with different topologies can be observed -- for example with one or two holes. Many of these dynamically transform into the most-stable planar compartment that has no holes.

\subsection{Increase in the average observable facet area}
\begin{figure}[htp]
\centering
\includegraphics[width=\columnwidth]{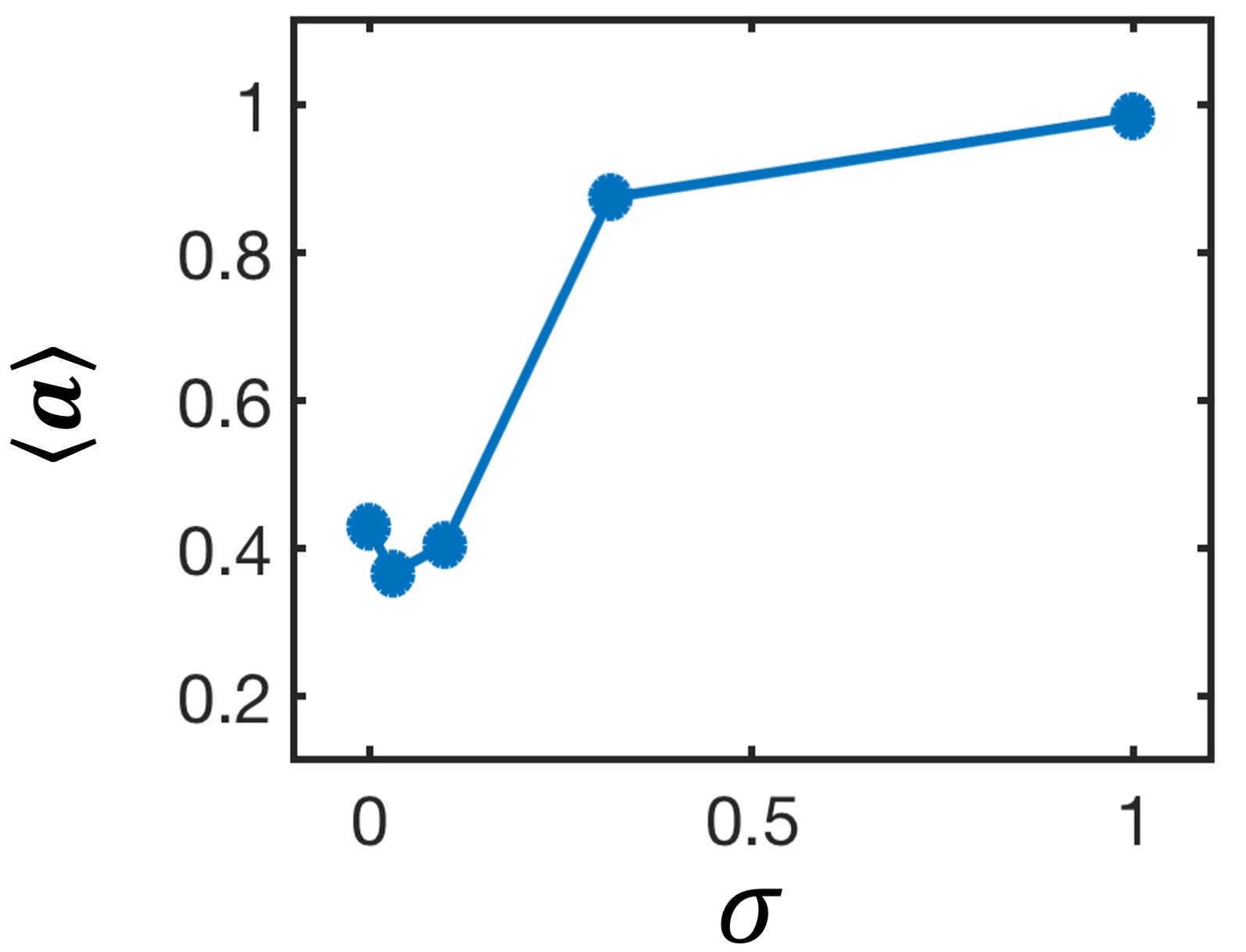}
\caption{(a) Plot for average observable interfacial facet area $\langle a \rangle $ as a function of tension $\sigma$. }
\label{fig:facet_max}
\end{figure}

We observe that with HIT, heterotypic facet areas become bimodal i.e. there are a majority of vanishingly small faces and the rest are larger faces. From an experimental point of view, vanishingly small facets are difficult to detect and hence, we plot the average of larger facets in Fig.~\ref{fig:facet_max}. We see that it increases directly with tension.

\subsection{Change in 2D registry leads to cusp-like response}
\label{2Dcalc}
\begin{figure}[htp]
\centering
\includegraphics[width=0.8\columnwidth]{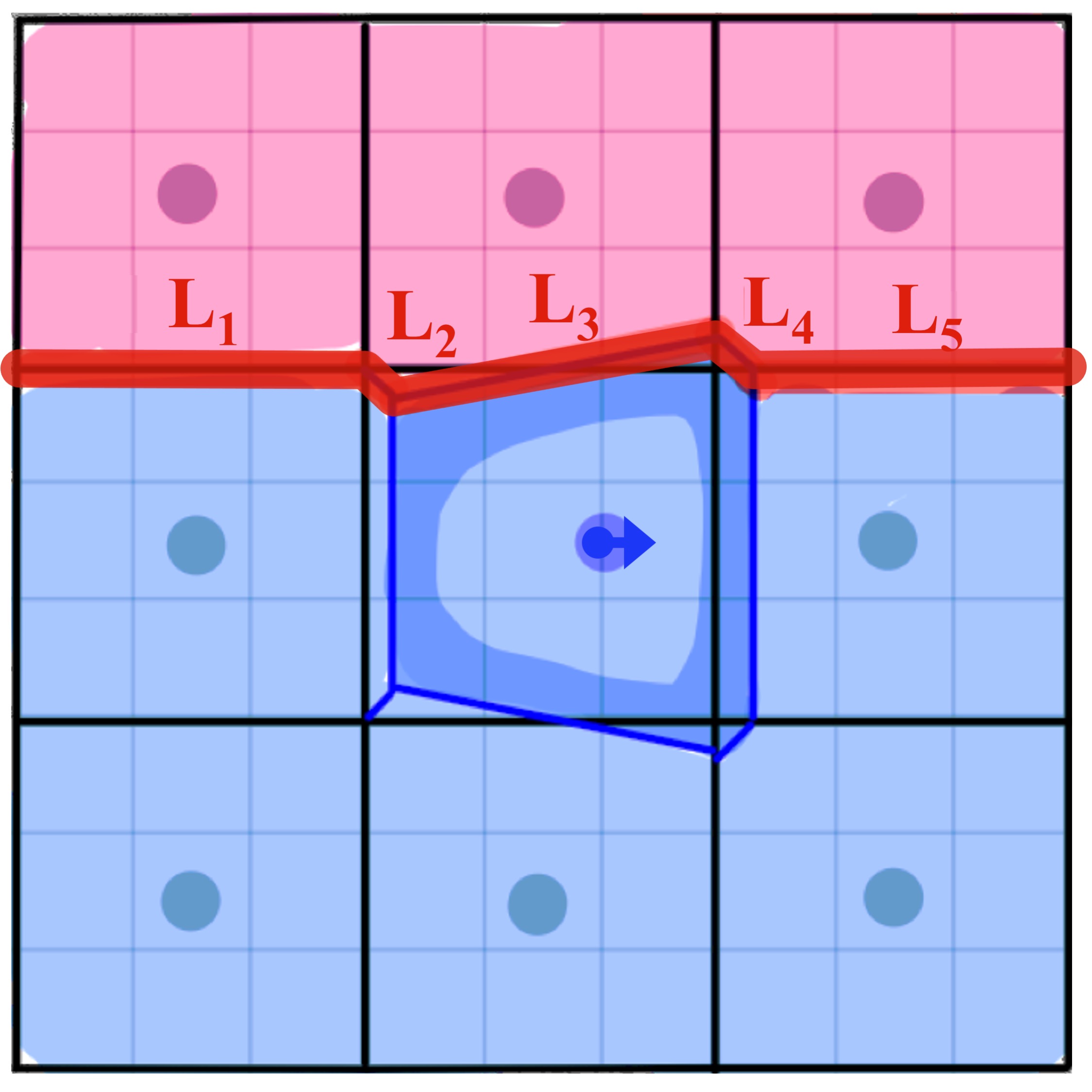}
\caption{A sketch of the 9-cell set up where the different cell types are colored with pink and blue and the heterotypic boundary highlighted in red. The Voronoi centers are shown in grey-filled circles . The black edges show the initial cell-cell boundaries. The central cell's boundary after the perturbation is highlighted in blue. }
\label{fig:9cell}
\end{figure}

For a perfect prism-like geometry along 3D interfaces, we need cells to be at the same distance from the interface as well as the cell pairs across the interface must have their centers aligned, i.e. registered. While the former condition is already understood using simple square lattice geometry calculation in a recent work~\cite{Sussman2018b}, the latter remains to be studied. Hence we use the exact same setup with 9 neighboring cells, but with a different local perturbation. As shown in Fig.~\ref{fig:9cell}, we consider a small perturbation of the central cell along the interface by an amount $\epsilon$, redraw the Voronoi tessellation and calculate final energy. 

The initial energy of the 9-cell system is:
\begin{equation}
 \label{eq:9cell_Ei1}
E_{i}=9k_A(a-a_0)^2+9(p-p_0)^2+3\gamma_0 l_{ij},
\end{equation}
where $\gamma_0$ is the interfacial tension, $l_{ij}$ is the length of edges shared between unlike cells. By making the assumptions of $s_0=1$ and $p_0=4$ to start from the ground state, we reduce the expression to:
\begin{equation}
 \label{eq:9cell_Ei2}
E_{i}=3\gamma_0,
\end{equation}
After displacing the central cell to the right, the new total energy to leading order becomes-
\begin{equation}
 \label{eq:9cell_Ef}
 \begin{aligned}
E_{f}=3\gamma_0+\gamma_0(\sqrt{2}-1)\epsilon_x\\
+\bigg(\frac{3k_A}{2}+(22-14\sqrt{2}+\gamma_0\big(\frac{20\sqrt{2}-8}{32\sqrt{2}}\big))\bigg)\epsilon_x^2,
\end{aligned}
\end{equation}

We see that to linear order, the energy for a perturbation parallel to the interface has exactly the same cuspiness (e.g. linear scaling with distance of perturbation $\epsilon_x$) that was observed for a perturbation perpendicular to the interface. Moreover, the coefficient is $\gamma_0\epsilon^2$, which is twice that of coefficient for a perpendicular perturbation~\cite{Sussman2018b}, indicating the non-linear stiffness is higher for a parallel (or sideways) perturbation.

\subsection{Change in 3D registry creates several new HIT facets}
\label{3Dcalc}

The calculation in the previous section establishes that a cell adjacent to a high-tension interface has a cusp-like response to both kinds of perturbations in 2D: perpendicular as well as parallel to the interface.
\begin{figure}[htp]
\centering
\includegraphics[width=0.9\columnwidth]{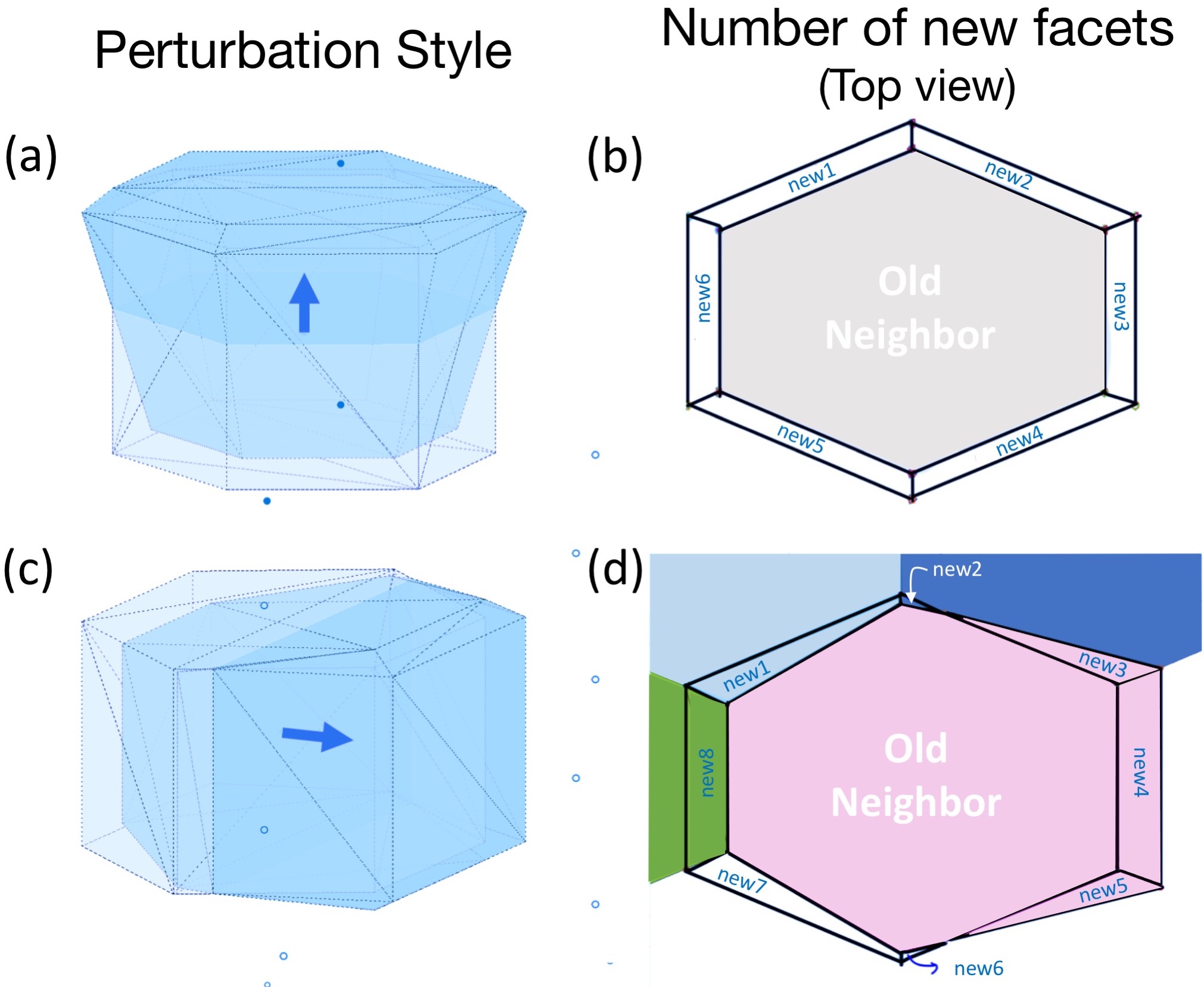}
\caption{A hexagonal right prism is perturbed both (a) perpendicular and (c) parallel to the shared horizontal interface at the top, as depicted by the blue arrows. The Voronoi cell centers are shown in blue circles. This gives rise to several new facets that can be seen from the top view on the right sub figures in (b) and (d). The original facet is labelled as `old neighbor' and the new facets are labelled as `new' and highlighted by dark edges. The perpendicular perturbation creates 6 new facets as shown in (b). The parallel perturbation creates 8 new facets as shown in (d). As the system is no longer symmetric about z-axis, three of the neighboring cells are colored differently to emphasize which cell the new facet belongs to. The perturbed cell is shown in pink. }
\label{fig:21cell}
\end{figure}

A qualitative similarity is only expected because the same number of new edges is created along the high tension cable for both kinds of perturbations. In order to extend this observation to 3D tissues one needs to check if new facets are being created along the high-tension interface, the same needs to be verified for a cell adjacent to a high-tension interface. To explore this we create a Voronoi tessellation of a group of hexagonal right prisms, stacked about their hexagonal faces. The isotropic hexagonal face of unit area has a 2D shape index of 3.72 and a the height of the prism is set to unity. This results in a 3D shape index of 5.72. In a manner very analogous to the 2D calculation performed in the previous section, we push a central prism both perpendicularly and parallely to the interface as shown in Fig.~\ref{fig:21cell}(a,c). The parallel perturbation shown here refers to perturbing the cell along the lateral direction that connects two neighboring centers.

As one would expect, the perpendicular perturbation leads to formation of 6 new facets as highlighted in the top view shown in Fig.~\ref{fig:21cell}(b). The parallel perturbation, on the other hand, leads to 8 new facets. Therefore, cusp-like behaviour is expected to accompany changes in registry in 3D systems in both cases. 

\subsection{BCC toy model}

\begin{figure}[htp]
\centering
\includegraphics[width=0.9\columnwidth]{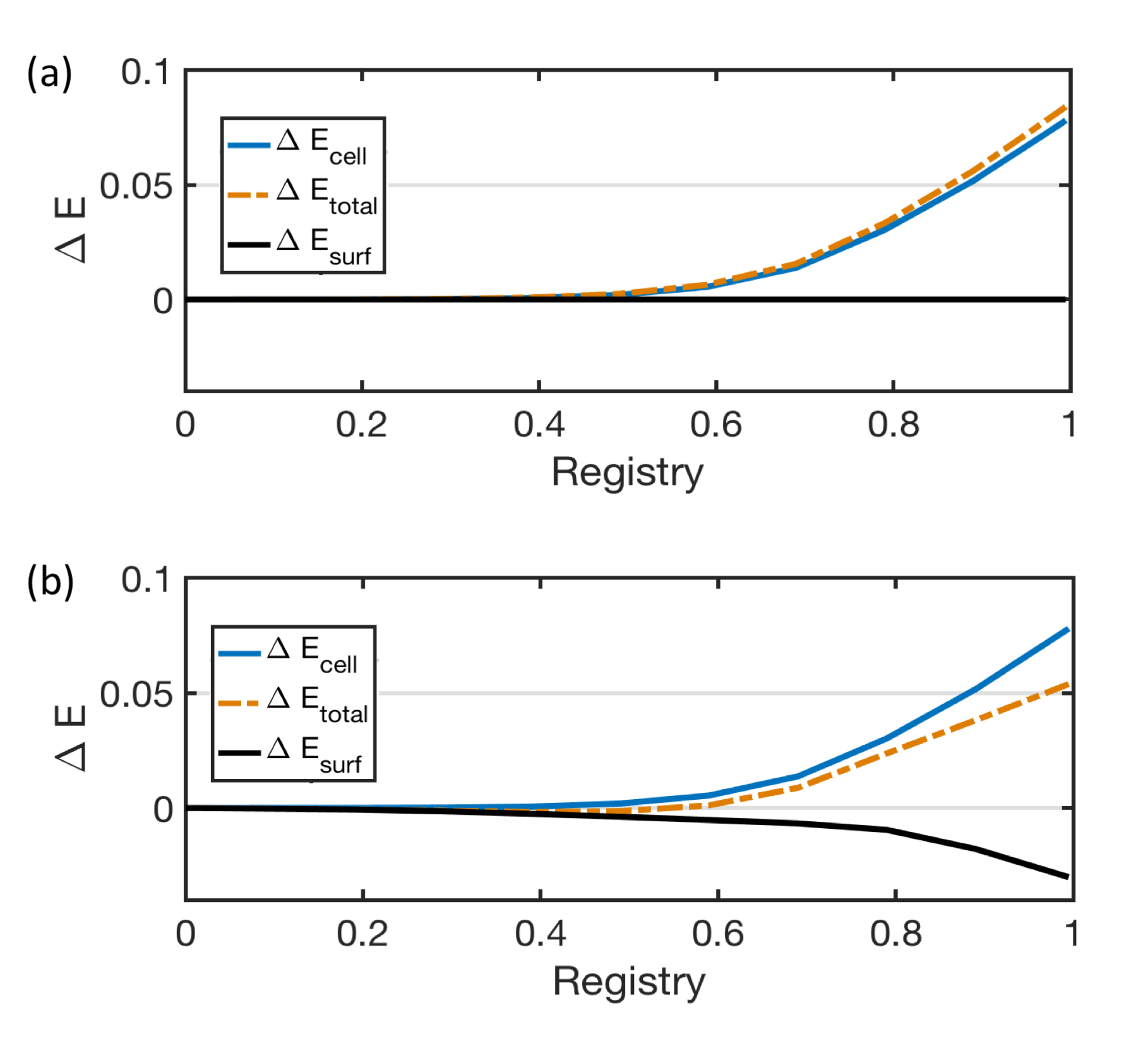}
\caption{ Shape versus surface energies at extreme tension values : Plot of change in total energy (in dashed orange), blue cell's mechanical energy (in solid blue) and interfacial energy (in solid black) with respect to registry is shown for (a) $\sigma=0.001$ and (b) $\sigma=1$. }
\label{fig:bcc_extra2}
\end{figure}

\begin{figure}[htp]
\centering
\includegraphics[width=\columnwidth]{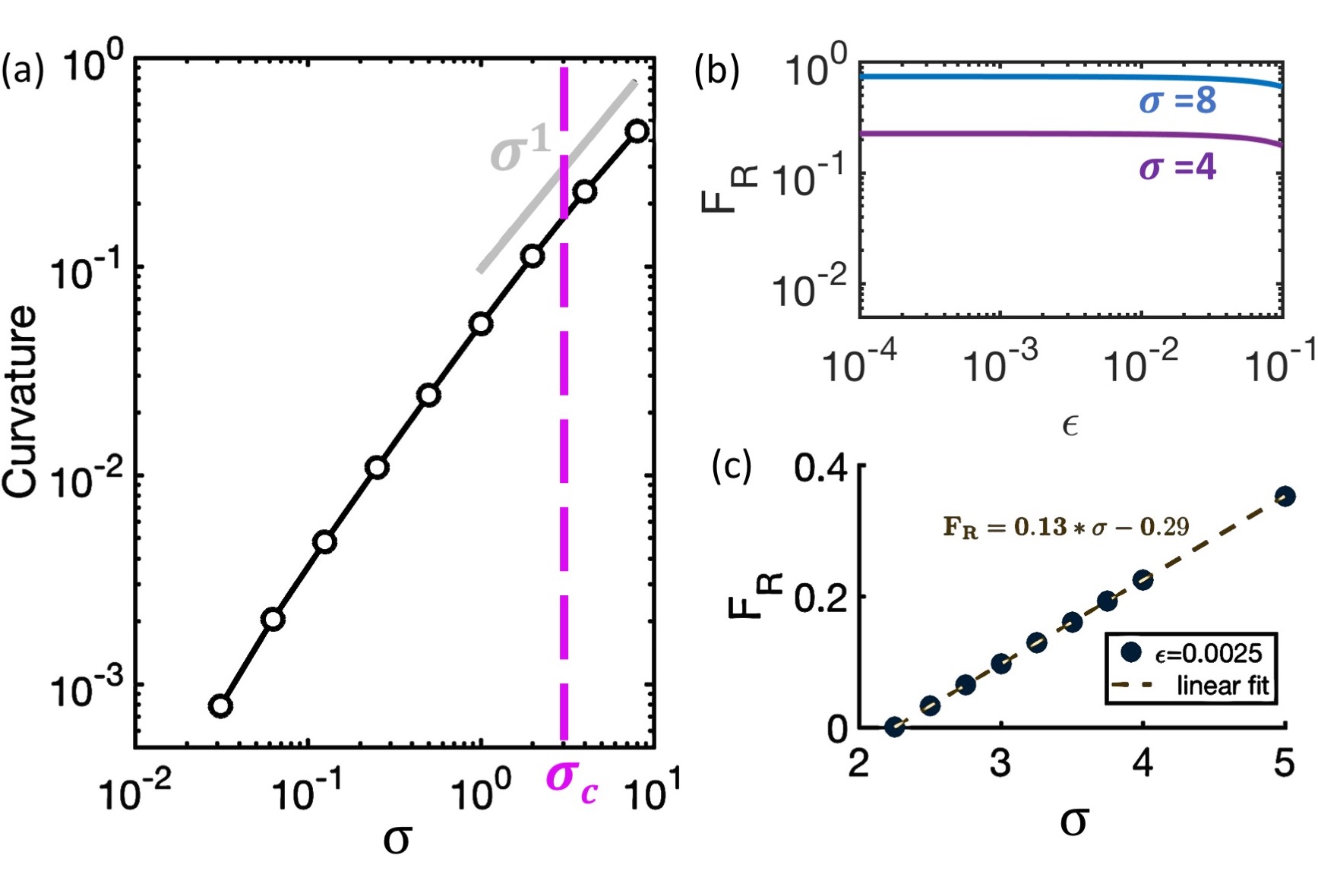}
\caption{Properties of parabolic and cuspy minima : (a) Plot of stiffness (magnitude of curvature) of the parabolic minima with respect to increasing tension. (b) The restoring force ($F_R$) for perturbations away from the cuspy minima is independent of the displacement on y axis for different values of tension ($\sigma$), except for the very largest displacements. (c) The value of restoring force at a fixed value of displacement is plotted as a function of tension, exhibiting a linear relationship.}
\label{fig:bcc_extra}
\end{figure}

We further explore the properties of both kinds of minima in Fig~\ref{fig:bcc_extra}. We also look at the contribution from shape frustration and surface energy to the total energy change of the system in Fig.~\ref{fig:bcc_extra2}.

\subsection{HCP toy model}
\begin{figure}[htp]
\centering
\includegraphics[width=\columnwidth]{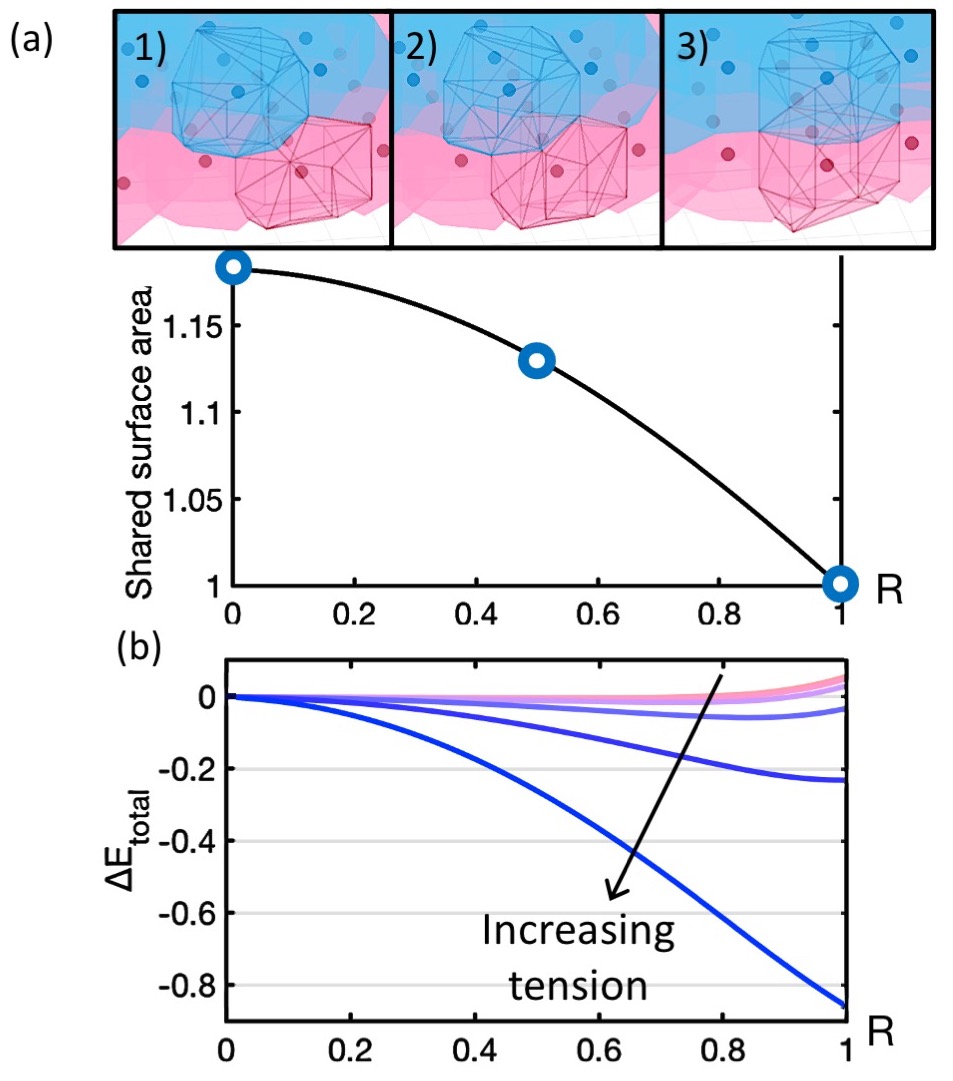}
\caption{ Global minima becomes registered for higher tension: (a) Shared surface area between both interfacial layers is computed as a function of the registration between the layers. The layer of blue cells is allowed to move past the string of layer of pink cells. Snapshots for no, half and complete registration is shown for encircled points. For better visualization, one cell from each layer is highlighted using solid-edge triangulation. (b) The change in the total energy of this system is plotted with respect to  registry, for different values of tension $\sigma=0.001,0.003,0.010,0.031,0.10,0.31$ shown in pink to blue, for a fixed shape value of 5.5. }
\label{fig:hcp_extra}
\end{figure}

For the HCP toy model, we explore the shared surface area and the system's energy profiles as a function of increased sigma. For shapes near transition, the behaviour is qualitatively very similar to the BCC toy model. The shared surface area is a minimum for complete registry as shown in Fig~\ref{fig:hcp_extra}(a). However, the ground state registry is unity only for very high tension values. For lower values, it depends on the cell shape. For a more fluid-like shapes the effect of shape frustration becomes negligible and the state prefers being registered for almost all values of tension. 

\subsection{Cusp-like restoring forces in bulk systems}

For the second part of this verification process i.e.,to check if the restoring forces are Hookean or non-Hookean in nature, one needs to analyse the qualitative scaling of the restoring forces with respect to perturbations about the ground state. This computation requires the fluid-like systems to be at mechanical equilibrium, though this is difficult in our simulations as we describe below. We let the steady state configurations found above to relax at zero fluctuations ($v_0=0$) using standard steepest descent dynamics. The integration time step is reduced to $dt=0.001$ and we integrate over a longer time window of $1 \times 10^6$ steps for a system-size $N=216$. With these choices, the noise floor, i.e. the magnitude of fluctuations in force that quantify how far the configuration is from the local minimum in the energy at the end of this many relaxation steps, reduces to a value of approximately $10^{-4}$. Unfortunately, we can not go below this value without major alterations to our code, because inhomogeneity in tensions in confluent models gives rise to higher fold coordinated vertices. In the current version of the 3D code, such configurations are not allowed and are thus computationally inaccessible. In order to decrease the number of such instances, we decrease the distance tolerance of the Voronoi algorithm by an order of magnitude to $10^{-12}$. Nevertheless, we still see a significant fraction of instances (e.g. for $\sigma=0.16$, $\sim65\%$ of the minimization procedures survive and for $\sigma=0.32$ $\sim20\%$ survive), and the number of such instances increases if we allow the system to relax for more time steps. Therefore, the choices for this minimization procedure have been chosen to balance between minimizing the number of many-fold-coordinated vertices while still having a small enough noise floor to be able to study truly minimized states. At the end of every successful minimization procedure, we then record the response of an interfacial cell to discrete changes in its cell-cell registry denoted by $\epsilon$.  

\begin{figure}[htp]
\centering
\includegraphics[width=1\columnwidth]{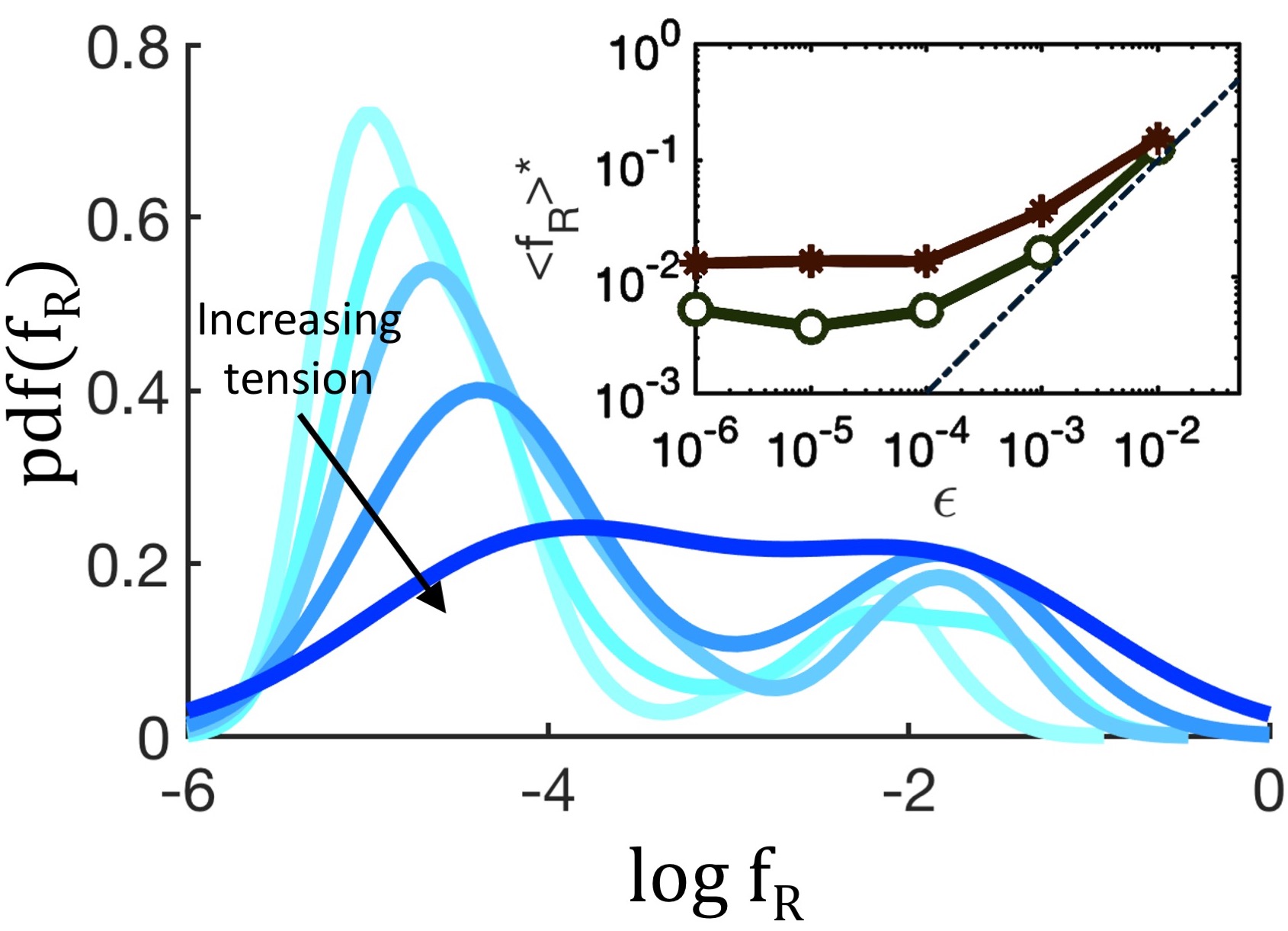}
\caption{ Cusp-like restoring forces observed at high tension : The probability distribution of restoring forces $f_R$ for the lowest $\epsilon=10^{-6}$ is shown as a function of tension values increasing from 0.16 (light blue) to 0.32 (deep blue) in steps of 0.04. In the inset, the mean value of $f_R$ for the second Gaussian (denoted by *) on the right is plotted for increasing $\epsilon$ for $\sigma=0.16$ in open circles and $\sigma=0.32$ in asterisk. The dashed line a Hookean fit to the large displacement data. }
\label{fig:SI_bulk}
\end{figure}

For the cases that survive, a randomly chosen interfacial cell is perturbed away from being registered by a discrete amount $\epsilon$ and we record the restoring force $f_R$ on the cell while keeping every other cell intact. The distribution of these forces is shown as a function of increasing tension in Fig.~\ref{fig:SI_bulk} for the lowest value of perturbation magnitude $\epsilon=10^{-6}$. Similar plots exist for higher values of $\epsilon$. For all cases, we observe a bimodal distribution of forces. The lower peak in the distribution corresponds to forces below the noise floor; in other words, these forces are consistent with systems that have not equilibrated even after $10^6$ time steps, and so our probe displacement of a single cell does not probe the energy minimum.  The higher peak is above the noise floor and represents real measurements of forces in response to perturbations away from a state that is truly in mechanical equilibrium. Therefore, we focus on the statistics of this higher-force peak in what follows. We expect that larger tensions are able to help equilibrate the system more effectively, shifting more configurations to the higher force bump. That is what is observed in our simulations: for $\sigma=0.16$, only $\sim15\%$ of the systems are on the right-hand-side Gaussian, while for $\sigma=0.32$, it rises to $\sim50\%$. The scaling of the average force in this higher peak (according to the best-fit Gaussian) is then shown as a function of perturbation magnitude $epsilon$. It is non-Hookean as  shown in the inset of Fig.~\ref{fig:SI_bulk}. Therefore, the computationally accessible ground states have a cusp-like behaviour.     

\end{document}